\documentclass[prb, floatfix, nobibnotes, reprint, superscriptaddress, hyperref]{revtex4-2}
\usepackage[utf8]{inputenc}
\usepackage{graphicx}
\usepackage{xcolor}
\usepackage{multirow}
\usepackage{booktabs}
\usepackage{amsmath}
\usepackage{mathtools}
\usepackage[version=4]{mhchem}

\usepackage{soul}

\begin{document}
\title{\papertitle}
\title{Learning to Converge: Warm-Starting DFTB Self-Consistent Charges with Machine Learning}

\author{Maximilian L. Ach}
\affiliation{Fritz-Haber-Institut der Max-Planck-Gesellschaft, Faradayweg 4-6, D-14195 Berlin, Germany}
\author{Karsten Reuter}
\affiliation{Fritz-Haber-Institut der Max-Planck-Gesellschaft, Faradayweg 4-6, D-14195 Berlin, Germany}
\author{Chiara Panosetti}
\email[Contact author: ]{panosetti@fhi.mpg.de}
\affiliation{Fritz-Haber-Institut der Max-Planck-Gesellschaft, Faradayweg 4-6, D-14195 Berlin, Germany}

\begin{abstract}

Semiempirical electronic structure methods such as Density-Functional Tight-Binding (DFTB) offer a computationally efficient approach to molecular and materials simulations, bridging the gap between first-principles accuracy and classical force field speed while retaining full access to electronic properties.
However, DFTB calculations based on self-consistent charge (SCC) schemes can still suffer from slow convergence, particularly for complex molecular and materials systems, making the iterative procedure a significant bottleneck in large-scale simulations and high-throughput workflows.
We present a machine learning approach that accelerates DFTB simulations by predicting optimal initial atomic charges.
Using element-specific models based on the Smooth Overlap of Atomic Positions descriptor and kernel ridge regression, we train charge models on reference calculations and demonstrate that ML-predicted initial charges consistently and significantly improve SCC convergence across diverse chemical systems including organic molecules, biomolecules, water clusters, transition metal oxides and solid electrolytes.

\end{abstract}


\maketitle

\section{Introduction}



Semiempirical methods are widely used in computational materials science and molecular simulation, because they offer quantum mechanical treatment of the electronic structure at a fraction of the cost of first-principles methods.\cite{thiel2014semiempirical, dral2024modern} 
Among these approaches, Density-Functional Tight-Binding (DFTB) has emerged as a particularly versatile framework\cite{porezag1995construction, seifert1996calculations, elstner1998self, oliveira2009density, grimme2017robust, bannwarth2019gfn2}, successfully applied to diverse systems ranging from organic and biological systems\cite{elstner2006scc, christensen2016semiempirical, morao2017rapid, schmitz2020quantum, kumar2023gpu} to solid-state materials, nanostructures, and catalytic interfaces.\cite{panosetti2021dftb, ricchebuono2023assessing, vuong2023toward, gupta2021using, chellam2025density, quaino2023dft, santos2024free}
The method's computational efficiency, typically 2-3 orders of magnitude faster than density functional theory (DFT), enables simulations of systems with thousands of atoms over nanosecond timescales, occupying an important niche between classical force fields and DFT.\cite{nishizawa2016three, allec2019heterogeneous, kumar2023gpu, gregory2022quantum, yagi2025high, spies2025structural}

Nowadays, machine-learned interatomic potentials (MLIPs) represent a complementary and increasingly important approach in this space.\cite{unke2021machine, jacobs2025practical}
Trained on reference data from DFT or higher-level methods, they enable faster simulations and make otherwise computationally prohibitive calculations feasible.\cite{butler2018machine, batatia2025foundation, riccius2025out}
Unlike DFTB, MLIPs lack direct access to the electronic structure and the full range of quantities it provides, and inherently non-local phenomena such as long-range electrostatic interactions or charge transfer remain challenging to describe, although notable progress has recently been made in this direction.\cite{ko2021fourth, staacke2021role, loche2025fast, vondrak2025pushing}

Previous work has also demonstrated the potential of applying machine learning (ML) techniques to DFTB simulations.
This includes the development of ML-optimized electronic parameters\cite{fan2022obtaining, sun2023machine, mcsloy2023tbmalt, fan2025advancing, xia2026nn} or directly learning the corresponding Slater-Koster integrals\cite{li2018density, mcsloy2023tbmalt, choudhary2025slakonet}, as well as the construction of ML-based repulsive potentials.\cite{kranz2018generalized, panosetti2020learning, stohr2020accurate, goldman2023enhancing} 
Here, instead of focussing on modeling the interactions, we demonstrate how ML can be leveraged to improve the convergence of self-consistent charges (SCC) in DFTB.

The SCC-DFTB approach---and, analogously, the extended Tight Binding (xTB), which shares much of the formalism with DFTB---requires iterative solution for atomic partial charges that reproduce the correct electrostatic interactions, particularly important for polar bonds, heterogeneous interfaces, and other systems with significant charge redistribution.
Despite its efficiency and versatility, DFTB and xTB calculations employing SCC frequently face slow convergence of the iterative charge optimization scheme.\cite{Elstner2007, Seabra2007, gaus2011dftb3, bannwarth2019gfn2, Bannwarth2020, hourahine2020dftb+, guibourg2024dftb, MahmoudiGahrouei2024, Pokora2025}
Poor charge initialization can lead to oscillatory behavior, possibly requiring hundreds to thousands of iterations to converge or failing entirely.
These challenges are particularly significant in large systems, complex chemical environments, or systems outside the scope of a given parameterization.
Experienced practitioners at times circumvent convergence issues by manually setting initial charges based on formal oxidation states, electronegativities or plain trial and error, but this approach is tedious, system-specific, and often inadequate for capturing the nuanced charge distributions in complex chemical environments.
Convergence issues not only waste computational resources but also limit the practicality of DFTB for high-throughput screening and automated workflows in materials discovery. 

Towards this end, machine learning can be used to predict accurate initial charges directly from atomic structures, reducing iteration counts and lowering convergence failure rates.
Learning SCC charges has previously been explored as a proof of concept for small charged silicon carbide clusters.\cite{guibourg2024dftb} 
In this approach, however, ML-inferred charges were used to bypass the SCC procedure entirely, followed by a single Hamiltonian diagonalization, yielding results that approximate full SCC-DFTB.

In this work, we employ ML-predicted charges as improved initial guesses while retaining the full SCC procedure, so that the resulting calculations remain fully converged.

The remainder of this paper is organized as follows: Section II describes the theoretical background of DFTB and the machine learning methodology.
Section III presents results for different chemical systems and compares the convergence behavior.
Section IV concludes with a discussion of limitations and future directions.



\section{Theory \& Methods}

\subsection{Density-Functional Tight-Binding}

The DFTB method approximates the Kohn-Sham DFT total energy through a second- or higher-order expansion of the DFT total energy functional with respect to charge density fluctuations.
In the Self-Consistent-Charge formulation (SCC-DFTB), the total energy of the system (up to second order) is expressed as:
\begin{equation}
E_{\text{SCC}} = \sum_{i}^{\text{occ}} f_i \langle \psi_i | \hat{H}^0 | \psi_i \rangle + \frac{1}{2} \sum_{I,J} \gamma_{IJ} \Delta q_I \Delta q_J + \sum_{I<J}V_{IJ}^{\text{rep}}
\label{eq:SCC_energy}
\end{equation}

where $f_i$ are occupation numbers, $\psi_i$ are Kohn-Sham orbitals, $\hat{H}^0$ is the reference Hamiltonian (typically derived from single-atom DFT), $\gamma_{IJ}$ describes the screened Coulomb interaction between Mulliken charge fluctuations $\Delta q_I$ and $\Delta q_J$ on atoms $I$ and $J$, and $V_{IJ}^{\text{rep}}$ represents pairwise repulsive interactions.

Charge transfer is introduced by expanding the DFT energy to second order around the single-atom reference density, i.e. $n(\mathbf{r}) = n_0(\mathbf{r}) + \delta n(\mathbf{r})$.
The electronic states are expanded in a minimal atomic orbital basis $\{\phi_\mu\}$, leading to the generalized eigenvalue problem:

\begin{equation}
\sum_\nu (H_{\mu\nu} - \varepsilon_a S_{\mu\nu}) c_{\nu a} = 0
\end{equation}

where $S_{\mu\nu}$ are overlap matrix elements, $\varepsilon_a$ are orbital energies, and $c_{\nu a}$ are molecular orbital coefficients.
The Hamiltonian matrix elements are charge-dependent and given by

\begin{equation}
H_{\mu\nu} = H_{\mu\nu}^0 + \frac{1}{2} S_{\mu\nu} \sum_K (\gamma_{IK} + \gamma_{JK}) \Delta q_K
\end{equation}

where $\mu$ and $\nu$ are centered on atoms $I$ and $J$ respectively, coupling the electronic structure to the charge distribution.\cite{koskinen2009density}

The partial charges ${q_I}$ are determined self-consistently through an iterative procedure that enforces stationarity of the (variational) DFTB energy functional from Eq. \ref{eq:SCC_energy}:

\begin{enumerate}
\item Initialize atomic charges (typically neutral)
\item Construct the Hamiltonian matrix using current charges
\item Solve Kohn–Sham–like secular equations for eigenvalues and orbitals
\item Calculate new Mulliken charges from the density matrix
\item Check convergence. If not converged, mix old and new charges and repeat from step 2
\end{enumerate}

Convergence is achieved when the change in charges between iterations falls below a threshold (for example $10^{-6}$ electrons).
However, poor initial guesses can lead to charge oscillations or divergence, requiring hundreds or thousands of iterations or causing complete failure.
Our approach aims to provide physically meaningful initial charges that position the SCC procedure closer to its minimum.

\subsection{SOAP Descriptors for Atomic Environments}

To predict atomic charges from the structure, we require a representation that captures the local chemical environment of each atom.
With straightforward transferability to other representations, we here employ the Smooth Overlap of Atomic Positions (SOAP) descriptor, which provides a rich description of atomic neighborhoods in a form that is invariant to rotations, translations, and permutations of identical atoms.\cite{soap2013}

For an atom at position $\mathbf{r}_0$, the SOAP descriptor is constructed from the atomic neighbor density:

\begin{equation}
\rho(\mathbf{r}) = \sum_{i \in \text{neighbors}} g(|\mathbf{r} - \mathbf{r}_i|, Z_i) f_c(|\mathbf{r}_i - \mathbf{r}_0|)
\end{equation}

where $g$ is a Gaussian smearing function associated with atom $i$ of element type $Z_i$, and $f_c$ is a smooth cutoff function that defines the extent of the local environment.
The neighbor density is expanded in a basis of radial functions $g_n(r)$ and spherical harmonics $Y_l^m(\theta, \phi)$:

\begin{equation}
\rho(\mathbf{r}) = \sum_{nlm} c_{nlm} g_n(r) Y_l^m(\theta, \phi)
\end{equation}

with radial quantum number $n \leq n_{\text{max}}$ and angular quantum number $l \leq l_{\text{max}}$ as hyperparameters, allowing for a high-resolution description of the local environment.

Rotational invariance is achieved by constructing the power spectrum:

\begin{equation}
p_{nn'll'}^{ZZ'} = \sum_{m=-l}^{l} c_{nlm}^{Z*} c_{n'l'm}^{Z'}
\end{equation}

The complete SOAP descriptor for an atom consists of the set $\{p_{nn'll'}^{ZZ'}\}$ for all relevant combinations of $n$, $n'$, $l$, $l'$, $Z$, and $Z'$.
This yields a fixed-length feature vector that effectively characterizes the local atomic environment while respecting relevant physical symmetries.

\subsection{Kernel Ridge Regression}

Given SOAP descriptors $\mathbf{x}_i$ and corresponding reference charges $q_i$ for atoms in our training set, we want to learn a regression function $f(\mathbf{x})$ that predicts charges for new atomic environments.
We employ kernel ridge regression (KRR), which predicts charges for a new environment $\mathbf{x}_*$ as:

\begin{equation}
f(\mathbf{x}_*) = \sum_{i=1}^{N_{\text{train}}} \alpha_i k(\mathbf{x}_*, \mathbf{x}_i)
\end{equation}

where $k(\mathbf{x}, \mathbf{x}')$ is a kernel function measuring similarity between environments, and the weights $\boldsymbol{\alpha}$ are obtained by solving:

\begin{equation}
\boldsymbol{\alpha} = (\mathbf{K} + \lambda \mathbf{I})^{-1} \mathbf{q}
\end{equation}

with kernel matrix $K_{ij} = k(\mathbf{x}_i, \mathbf{x}_j)$, regularization parameter $\lambda$, and training charges $\mathbf{q}$.

For the kernel function, we employ the radial basis function (RBF) kernel:

\begin{equation}
k_{\text{RBF}}(\mathbf{x}, \mathbf{x}') = \exp\left(-\frac{\|\mathbf{x} - \mathbf{x}'\|^2}{2\sigma^2}\right)
\end{equation}

with length-scale parameter $\sigma$.
A key design principle of our approach is \textit{element-specific modeling}: we train separate regression models for each chemical element rather than a single global model.
This improves prediction accuracy by allowing each model to specialize to element-specific charge patterns, while simultaneously reducing model size and computational cost compared to a single multi-element model.
Additionally, element-specific models enhance transferability across chemical domains, as models for common elements like carbon and hydrogen can be reused across different systems.
At inference time, charges for a new molecular structure are predicted by applying the appropriate element-specific model to each atom and combining the predictions.




\subsection{Datasets and Reference Charges}
\label{sec:Datasets}


We evaluate our method on a diverse set of chemical systems.
Besides using converged DFTB Mulliken charges as training targets for our charge models, we also investigate learning DFT-level charges.
The primary and most natural choice is to train on aforementioned DFTB charges ideally obtained from the same parameterization as used at inference time.
Since the ML model learns to reproduce the DFTB charge distribution directly, the predicted initial charges are well-matched to the SCC target and are expected to yield the largest convergence improvements.
The drawback is that dedicated DFTB reference calculations must be performed for each chemical system and parameter set of interest.
An alternative is to leverage the vastly larger collection of publicly available DFT data, which enables charge models to be trained without any DFTB reference calculations.
The inherent shortcoming is that DFT and DFTB charge populations differ systematically, owing to differences in basis sets, exchange-correlation treatment, and the tight-binding approximations.

For each dataset, we recompute SCC-DFTB charges using DFTB$+$ \cite{hourahine2020dftb+, hourahine2025recent} (version 24.1) to obtain converged Mulliken charges for all structures.
Unless stated otherwise, all calculations employ the second-order (DFTB2) formulation.
Additional details on the DFTB calculations can be found in Sec.~S1 of the SM.


\textbf{QM9.}
The QM9 dataset\cite{ramakrishnan2014quantum} contains nearly 134,000 small organic molecules (C, H, N, O, F) with up to 9 heavy atoms.
We compute DFTB Mulliken charges using the 3ob-3-1 (hereafter referred to as 3ob) parameter set.\cite{gaus2013parametrization, kubillus2015parameterization} 
The original dataset also contains B3LYP/6-31G(2df,p)\cite{becke1993density, lee1988development, ditchfield1971self, frisch1984self, hehre1972self, krishnan1980self} Mulliken charges.

\textbf{Ni$_x$O$_y$.}
The Ni$_x$O$_y$ dataset comprises 20,858 structures spanning metallic nickel and a range of oxide stoichiometries---including \ce{NiO}, \ce{Ni2O3}, and more oxygen-rich phases up to \ce{NiO4}---capturing a broad range of Ni and O coordination environments relevant to heterogeneous catalysis.
It was generated using a genetic algorithm for Ni–O binary bulk systems and contains up to 16 atoms per structure.\cite{song2026adaptive}
DFTB Mulliken charges are computed using the mio-1-1 (hereafter referred to as mio) parameter set\cite{elstner1998self} for oxygen, as well as its trans3d extension \cite{zheng2007parameter} for nickel and nickel-oxygen interactions.
Additionally, we calculated DFT Mulliken charges at the PBE level\cite{perdew1996generalized} using FHI-aims (version aims.260331) with the \textit{light} numerical atom-centered orbital basis set.\cite{blum2009ab}
The calculations employed automatic $k$-point grids with a spacing of 0.4\,\AA$^{-1}$, scalar-relativistic treatment via the atomic ZORA approximation, and non-spin-polarized settings.

\textbf{LLZO.}
\ce{Li7La3Zr2O12} (LLZO) is a garnet-type ceramic solid electrolyte of interest for all-solid-state lithium-ion batteries.
The LLZO dataset of Holland et al.\cite{holland2023workflow} provides over 2.1 million symmetrically unique 192-atom bulk structures of cubic LLZO, generated by systematically enumerating Li site occupancies.
For a subset of 1,235 structures, the dataset contains DFT data calculated with the ONETEP linear-scaling DFT code\cite{skylaris2005introducing, prentice2020onetep}, at the PBE level\cite{perdew1996generalized} with an optimized NGWF basis.\cite{skylaris2001accurate, skylaris2002nonorthogonal}
We draw a random subset of 5,000 structures and compute DFTB Mulliken charges using the PTBP parameter set.\cite{cui2024obtaining} 

\textbf{SPICE subsets.}
We use three subsets of the SPICE dataset\cite{eastman2023spice, eastman2024nutmeg} and compute DFTB Mulliken charges with the 3ob parameter set.\cite{gaus2013parametrization, gaus2014parameterization, kubillus2015parameterization}
Structures with a non-zero total charge are excluded, as we target neutral systems in this work.
The dataset additionally reports DFT-derived MBIS charges\cite{verstraelen2016minimal} at the $\omega$B97M-D3(BJ)/def2-TZVPPD level of theory.\cite{najibi2018nonlocal, mardirossian2016omegab97m, weigend2005balanced, rappoport2010property}
The selected SPICE subsets are:
\begin{itemize}
\item \textit{Water clusters}: 1,000 structures of 30-molecule clusters (90 atoms each).
\item \textit{Solvated amino acids}: 1,300 structures (1,000 of which neutral) of all 20 amino acids in explicit water (79--96 atoms, H/C/N/O/S).
\item \textit{Dipeptides}: 33,850 structures (20,950 of which neutral) covering all 400 pairwise combinations of the 20 amino acids (26–-60 atoms, H/C/N/O/S).
\end{itemize}

\section{Results}

\subsection{Charge Prediction Accuracy}

\begin{figure*}[htbp]
\centering
\includegraphics[width=\textwidth]{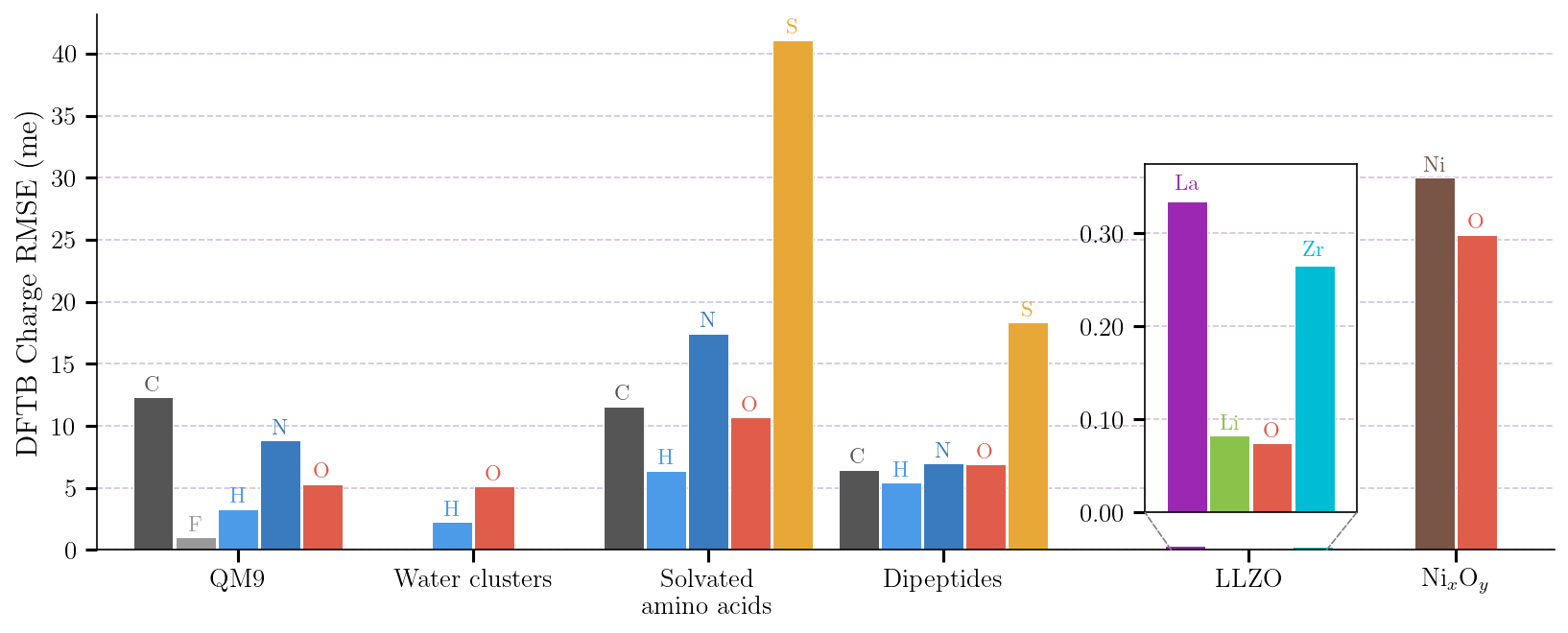}
\caption{DFTB charge prediction RMSE (in units of $10^{-3}$e) for all elements across datasets.}
\label{fig:rmse_barchart}
\end{figure*}

We train machine learning models for DFTB charge prediction across the six datasets presented in Sec. \ref{sec:Datasets}.
For each dataset, element-specific KRR charge models are trained on DFTB reference charges, with SOAP descriptors generated using the DScribe library.\cite{himanen2020dscribe, laakso2023updates}
Details on the reference data calculations and model training and hyperparameter selection are provided in Secs.~S1 and~S2 of the SM.
Figure~\ref{fig:rmse_barchart} shows the root mean squared error (RMSE) of the charge predictions on the test set for all datasets and elements (the corresponding MAE results are provided in Fig.~S1 of the SM).

Across all datasets, our models achieve low charge prediction errors, with accuracy varying depending on chemical diversity and training set size.
The bulk LLZO dataset yields particularly low errors, since the \ce{LaZrO} host lattice is fixed across all structures and only the Li site occupancy varies.\cite{holland2023workflow}
Therefore all atoms occupy well-defined crystallographic sites within a periodic, compositionally uniform lattice, limiting the range of distinct local environments.
In contrast, the SPICE biomolecular subsets are more challenging due to the larger chemical diversity, and training data being limited for elements such as sulfur.
Learning curves in Sec.~S4 of the SM demonstrate well-behaved improvement with training set size, indicating that errors would further decrease with more data.

\begin{table*}[htbp]
    \centering
    \caption{SCC convergence performance for different charge initialization strategies across all datasets, evaluated on held-out test sets using five-fold cross-validation.
    Reduction and fewer/equal/more percentages are relative to default zero-charge initialization.
    Gasteiger charges are only applied to the non-periodic systems.}
    \label{tab:scc_convergence}
    \begin{tabular*}{\textwidth}{@{\extracolsep{\fill}}clccccc}
\toprule
Dataset & Initialization & SCC Cycles & Reduction (\%) & Fewer Cycles (\%) & Equal Cycles (\%) & More Cycles (\%) \\
\midrule
\multirow{4}{*}{QM9}
& Default  & 10.2 & ---          & ---          & ---          & ---          \\ 
& DFTB-ML       & 7.3 & 28.3 & 99.8 & 0.2 & 0.0 \\ 
& Gasteiger     & 9.2 & 9.2 & 69.1 & 29.0 & 1.9 \\ 
& DFT-ML   & 9.2 & 9.5 & 70.7 & 28.1 & 1.3 \\ 
\midrule
\multirow{4}{*}{Water clusters}
& Default  & 9.3 & ---          & ---          & ---          & ---          \\ 
& DFTB-ML       & 6.2 & 33.6 & 100.0 & 0.0 & 0.0 \\ 
& Gasteiger     & 8.0 & 13.8 & 97.5 & 2.5 & 0.0 \\ 
& DFT-ML   & 8.1 & 12.8 & 91.0 & 9.0 & 0.0 \\ 
\midrule
\multirow{4}{*}{Solvated amino acids}
& Default  & 11.4 & ---          & ---          & ---          & ---          \\ 
& DFTB-ML       & 8.9 & 22.1 & 100.0 & 0.0 & 0.0 \\ 
& Gasteiger     & 10.6 & 7.3 & 73.0 & 27.0 & 0.0 \\ 
& DFT-ML   & 10.5 & 8.0 & 83.0 & 17.0 & 0.0 \\ 
\midrule
\multirow{4}{*}{Dipeptides}
& Default  & 11.9 & ---          & ---          & ---          & ---          \\ 
& DFTB-ML       & 8.2 & 31.5 & 100.0 & 0.0 & 0.0 \\ 
& Gasteiger     & 11.3 & 5.2 & 49.3 & 49.0 & 1.7 \\ 
& DFT-ML   & 10.3 & 13.8 & 99.0 & 1.0 & 0.0 \\ 
\midrule
\multirow{4}{*}{LLZO}
& Default  & 18.8 & ---          & ---          & ---          & ---          \\ 
& DFTB-ML       & 8.0 & 57.6 & 100.0 & 0.0 & 0.0 \\ 
& Gasteiger     & --- & --- & --- & --- & --- \\
& DFT-ML   & 17.0 & 9.3 & 99.3 & 0.7 & 0.0 \\ 
\midrule
\multirow{4}{*}{Ni$_x$O$_y$}
& Default  & 62.5 & ---          & ---          & ---          & ---          \\ 
& DFTB-ML       & 10.0 & 84.0 & 99.5 & 0.3 & 0.3 \\ 
& Gasteiger     & --- & --- & --- & --- & --- \\
& DFT-ML   & 23.2 & 62.9 & 86.1 & 5.6 & 8.3 \\ 
\bottomrule
\end{tabular*}
\end{table*}

\begin{figure*}[htbp]
    \centering
    \includegraphics[width=\textwidth]{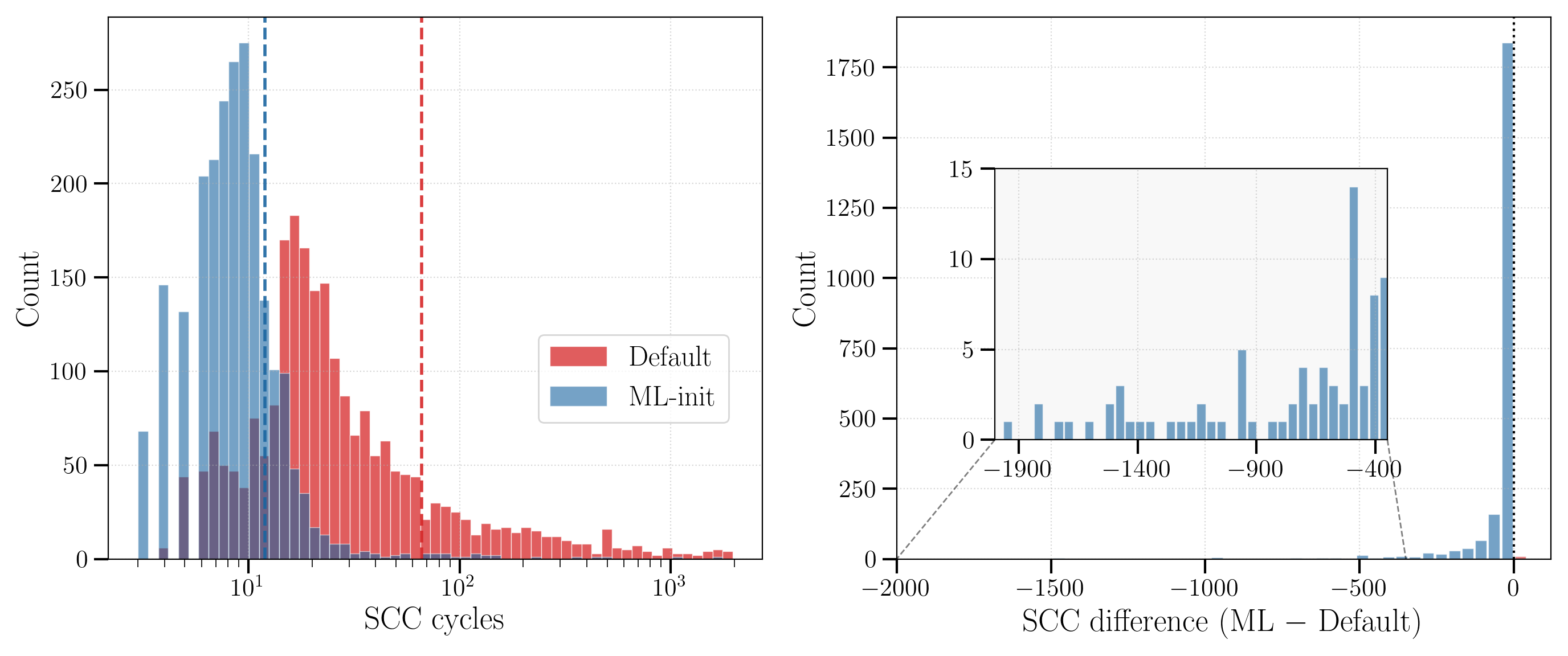}
    \caption{SCC iteration count distributions for Ni$_x$O$_y$ with default and ML-predicted (ML-init) charge initialization (left), and the per-structure cycle difference ML-init minus Default (right). The inset highlights structures with the largest reductions.}
    \label{fig:scc_nio}
\end{figure*}

\subsection{Impact on DFTB SCC Convergence}

After training our DFTB charge models, we apply them to out-of-sample test structures by predicting initial charges for SCC-DFTB.
We evaluate the impact on SCC iteration count and compare our ML-predicted charges to the DFTB$+$ default of initializing all atomic charges to zero (neutral atoms).
Table~\ref{tab:scc_convergence} summarizes our results across all datasets.
For the non-periodic systems, we additionally include Gasteiger charges\cite{gasteiger1980iterative} as a simple reference baseline.
We also tested the classical Rappe-Goddard charge equilibration (QEq)\cite{rappe1991charge}, but found it unreliable as an initial guess, since in several cases the charges fell outside the physically acceptable range and prevented calculations from starting.
If applicable, training sets were capped at 60,000 atomic environments per element model to limit kernel size.
Model hyperparameters were selected by five-fold cross-validation and all reported metrics are evaluated on a separate held-out test set (more details in Sec.~S2 of the SM).
We employ the DFTB$+$ default charge convergence threshold of $10^{-5}$\,e and set the maximum number of SCC cycles to 2,000.
Results obtained using a stricter $10^{-6}$\,e threshold are provided in Table~S3 of the SM.
They support the same conclusions, though as expected the relative improvement compared to the total number of SCC cycles slightly decreases with a tighter convergence criterion.

Across all datasets, ML-predicted charges substantially reduce SCC iteration counts and outperform default initialization for nearly all test structures. 
Since each iteration requires the same full Hamiltonian construction and diagonalization, the iteration count is directly linked to computational cost, and these reductions translate directly to shorter wall times.
The magnitude of improvement is largely governed by the system and the degree of charge fluctuation within it.
The largest gains occur for Ni$_x$O$_y$ and LLZO, where equilibrium charges deviate substantially from neutral, making zero-charge initialization particularly inadequate.
In comparison, the molecular datasets show more modest, yet still significant and very consistent speedups.
In particular, for Ni$_x$O$_y$, the default initialization produces a broad distribution extending to several thousand cycles for a non-negligible fraction of structures, whereas ML initialization compresses these to a narrower distribution well below 20 cycles yielding an overall reduction of nearly 84\% in number of SCC cycles (Fig.~\ref{fig:scc_nio}).
This reflects the inherent difficulty of zero-charge initialization for transition metal oxides, where charge transfer between Ni and O is large and strongly dependent on local coordination.
LLZO benefits most from the regularity of its garnet crystal environment, which allows the model to predict charges near the converged solution, reducing mean cycle counts by 58\%.
For the three SPICE subsets, cycle counts are reduced by 34\% for water clusters, 32\% for dipeptides, and 22\% for solvated amino acids.
The more modest improvement for solvated amino acids reflects the lower charge model accuracy (see Figure~\ref{fig:rmse_barchart}), particularly for sulfur, which appears in only two amino acids and thus provides limited training examples.
Dipeptides perform better owing to the larger training set spanning all 400 amino acid pair combinations, though sulfur remains comparatively less accurate than other elements.
As previously mentioned, we found the learning curves to be well-behaved, which suggests that further cycle count reductions are readily achievable by increasing the training data.

While each DFTB parameterization in principle produces its own charge distribution, we still find that models trained on one parameterization consistently improve SCC convergence when applied to calculations with a different parameterization.
We tested 3ob, mio, and PTBP charge models against test sets with all three parameterizations (yielding nine combinations) across QM9 and the three SPICE subsets (see Table~S5 in the SM).
In every case, using a charge model trained on a different parameterization still outperforms the zero-charge default.
In general, transferability is linked to the similarity between parameterizations, since closely related parameter sets produce more closely aligned charge distributions.
Transfer between 3ob and mio is particularly effective, since 3ob was developed as a third-order extension of mio, whereas PTBP differs more substantially from both.
This likely reflects the fact that the PTBP parameterization was developed as a periodic table baseline parameter set and produces charge distributions that differ more systematically from both 3ob and mio across organic and biochemical systems.
Overall, cross-parameterization transferability reduces the need to generate dedicated reference calculations for every parameter set of interest, though the benefit is largest when the source and target parameterizations share a similar physical treatment of charge.
Gasteiger charges, as a simple topology-based method requiring no training data, also provide a somewhat consistent, yet considerably smaller improvement.

Beyond reducing the number of SCC cycles for structures that converge under both schemes, ML-predicted charge initialization can also improve the rate of convergence.
Convergence failure can occur because the SCC converges too slowly to finish within the allowed iterations, or because the initial guess places the calculation in a region from which a minimum cannot be reached.
To this end, we rerun SCC-DFTB using ML-predicted initial charges for structures that did not converge under default initialization and summarize the results in Table~\ref{tab:convergence_failures}.
Of the 1,118 previously unconverged Ni$_x$O$_y$ structures, 891 now converge, as well as 2 of the 23 dipeptide structures.
For QM9, all 3 structures that failed under default initialization also fail using ML initialization.
The structures that remain unconverged likely have intrinsic convergence difficulties that cannot be resolved by improving the initial guess alone, though for Ni$_x$O$_y$, where failures were most prevalent, the majority of previously failing structures converge after ML-predicted initialization.
In order to distinguish between structures that merely converge very slowly and those that fail to find a minimum altogether, we further extended the maximum SCC cycle value to 5,000 for the previously unconverged Ni$_x$O$_y$ structures.
As shown in Table~\ref{tab:convergence_failures}, an additional 166 structures converge under default initialization (952 remaining), while under ML initialization 942 of the 1,118 structures now converge (176 remaining).
\begin{table}[htbp]
\centering
\caption{Structures failing to converge under default initialization, and how many of these still fail after re-running with ML-predicted initial charges, evaluated at maximum SCC cycle limits of 2,000 and 5,000.}
\label{tab:convergence_failures}
\begin{tabular*}{\columnwidth}{@{\extracolsep{\fill}}lcccc}
\addlinespace[2pt]
\toprule
\addlinespace[4pt]
& \multicolumn{2}{c}{Default-init} & \multicolumn{2}{c}{ML-init} \\
\cmidrule(lr){2-3} \cmidrule(lr){4-5}
Dataset \hfill\ $|$ \ Max SCC & 2,000 & 5,000 & 2,000 & 5,000 \\
\midrule
QM9 & 3 & 3 & 3 & 3 \\
Water clusters & 0 & --- & 0 & --- \\
Solvated amino acids & 0 & --- & 0 & --- \\
Dipeptides & 23 & 23 & 21 & 21 \\
LLZO & 0 & --- & 0 & --- \\
Ni$_x$O$_y$ & 1,118 & 952 & 227 & 176 \\
\bottomrule
\end{tabular*}
\end{table}
The substantially larger gain under ML initialization demonstrates that the improved starting point not only accelerates convergence, but also enables many structures to reach the SCC minimum that otherwise remains inaccessible from the zero-charge starting point (within a practical SCC cycle budget).

We also recomputed the datasets and trained charge models using DFTB3\cite{gaus2011dftb3} for QM9 and the SPICE subsets, and find comparable or even slightly larger SCC cycle reductions than for DFTB2 (see Table~S4 in the SM).
This was not done for Ni$_x$O$_y$ and LLZO, since the mio, trans3d and PTBP parameters do not support DFTB3, however we expect the same result trends for these systems.


A limitation of the current approach is that ML models predict atom-resolved charges, without distinguishing between the contributions of individual shells (e.g.\ the s, p, and d occupations of a transition metal atom).
In a non-shell-resolved SCC calculation, DFTB$+$ distributes a provided initial atom charge across the shells of each atom by filling them sequentially in the order defined by the Slater-Koster parametrization, filling each shell to its maximum capacity before populating the next.
For elements with partially filled d-shells such as Ni, this distribution may not be physically consistent, potentially introducing an unfavorable starting point for the SCC even when the atom-resolved initial charges are close to the converged values.
This manifests in rare edge cases where ML-predicted initial charges lead to slightly slower convergence compared to zero-charge initialization, accounting for the 0.3\% of Ni$_x$O$_y$ test structures with increased cycle counts (Table~\ref{tab:scc_convergence}).
Extending the approach to shell-resolved charge models, i.e. predicting separate occupations for each angular momentum channel, would provide a more faithful and physically consistent initialization and represents a natural direction for future work.

\subsection{DFT Charges as Initial Guesses}

DFT calculations are more expensive than DFTB, yet vastly more DFT data is publicly available.
It is therefore natural to evaluate whether pre-existing DFT charges can serve as useful initial guesses without requiring dedicated DFTB reference calculations.
Therefore, we assess whether DFT-level charge models can also be used to improve SCC-DFTB convergence.
For QM9, LLZO, and Ni$_x$O$_y$, we train DFT Mulliken charge models, while for the SPICE subsets we use MBIS\cite{verstraelen2016minimal} charges, as outlined in Sec.~\ref{sec:Datasets}.

The DFT-initialized calculations consistently outperform the zero-charge default, but as expected, by a considerably smaller margin than the DFTB-trained ML models (Table~\ref{tab:scc_convergence}).
Reductions are modest and broadly consistent across all molecular datasets: roughly 10\% for QM9 and 8--14\% for the SPICE subsets, with dipeptides and water clusters responding somewhat better than solvated amino acids, and LLZO falling in the same range at 9\%.
The exception is Ni$_x$O$_y$, where DFT initialization achieves a 63\% reduction, reflecting the large charge deviations from neutrality and the comparatively good agreement between DFT and DFTB charges.
The gap is expected, since even though DFT and DFTB share the same physical picture, their charge populations often differ systematically, meaning that DFT charge models introduce a residual initial error that DFTB-trained models do not contain.
Parity plots comparing DFT and DFTB charge populations for each dataset are provided in Sec.~S7 of the SM.
The agreement between the two naturally determines how useful DFT-trained models can be as initial guesses.
Nonetheless, the result demonstrates that reference charges obtained outside of DFTB often provide better starting points than the zero-charge default and provide a useful baseline when no DFTB reference calculations are available (yet).

\section{Conclusions}
We have presented a machine learning approach to accelerate DFTB self-consistent charge calculations by providing improved initial charges prior to the SCC procedure.
Using element-specific KRR models with SOAP descriptors, we trained charge models on converged DFTB Mulliken charges and demonstrated their effectiveness across a broad range of chemical systems spanning organic molecules, biomolecules, water clusters, metallic and oxidized nickel, and the solid electrolyte \ce{Li7La3Zr2O12} (LLZO).
Across all six datasets, ML-predicted initial charges consistently and substantially reduce SCC iteration counts compared to the standard zero-charge initialization, requiring fewer cycles than the default in 99--100\% of test structures.
These gains translate directly to shorter wall times, since each SCC iteration carries essentially the same computational cost.
We find that the magnitude of improvement scales with the degree of charge redistribution in the system, as well as the accuracy of our charge models.
Cycle count reductions range from roughly 22\% for solvated amino acids, where limited training data constrains the accuracy of the initial charges, up to 84\% for Ni$_x$O$_y$, where large charge transfer makes neutral initialization particularly inadequate.
For LLZO, where the periodic garnet environment is highly regular, SCC iterations are reduced by 58\%.

Besides accelerating convergent calculations, the ML initialization is in some cases also capable of recovering structures, which previously did not converge.
We note, moreover, that the charge models presented here only represent a starting point.
Since model accuracy directly translates to better initialization and greater convergence gains, there is significant headroom for improvement and the reported results may be viewed as a lower bound on what is achievable.
Indeed, a more fundamental extension of the current approach would be to predict shell-resolved charges, providing separate initial occupations for each angular momentum channel rather than a single atom-resolved value.
This is for instance particularly relevant for transition metals with partially filled d-shells, where the current approach can in rare cases introduce an unfavorable shell-occupation, even when the total atomic charge is accurate.

We additionally showed that charge models trained on one DFTB parameterization still improve SCC convergence when applied to calculations using a different parameterization.
Transferability is best between closely related parameter sets such as 3ob and mio, but even mismatched models consistently outperform zero-charge initialization.

In cases, where no DFTB reference data is available (yet), models trained on DFT-level charges may provide a practical fallback.
Although the improvement is noticeably smaller than with DFTB-trained models - reflecting the systematic differences between DFT and DFTB charge populations - DFT-based initial charges still consistently outperform the zero-charge default across all datasets.
The approach is most effective when the two charge populations align well, and although this can be difficult to anticipate a priori, in practice we found that even DFT charges from different partitioning schemes consistently improved SCC-DFTB convergence.

Further, even simple non-ML charge schemes such as Gasteiger charges can deliver modest improvements without requiring any training data, although the gains are generally smaller than those of the ML-based approaches.

Beyond the quantitative improvements, our approach also automates charge initialization, removing the need for the manual, system-specific trial and error that experienced practitioners at times resort to when convergence is poor.
It is most impactful for high-throughput screening and automated simulation pipelines, where large numbers of structures are (ideally) processed without manual intervention.
Another area of application is the fitting of DFTB parameters, where a (very) large number of SCC calculations needs to be performed across chemically diverse structures.
For molecular dynamics and geometry optimization, the benefits are more limited, since charges from the previous step can serve as a warm start, making ML initialization merely useful at trajectory starts or restarts.

Future work could explore larger, more transferable models trained across multiple chemical databases, or extend the approach to other DFTB-based methods such as GFN-xTB and GFN2-xTB.
While kernel ridge regression is well-suited for interpolation within a distribution, neural network architectures may further offer advantages both in capturing longe-range effects and in transferability across chemical domains.

In our view, this work represents an effort complementary to machine-learned interatomic potentials, integrating machine learning into the electronic structure simulation workflow rather than using it as a surrogate model for the latter.

\section*{Author Contributions}

\noindent \textbf{Maximilian L. Ach}: Conceptualization; Data Curation; Formal analysis; Investigation, Methodology; Software; Visualization; Writing - Original Draft; Writing - Review \& Editing.
\textbf{Karsten Reuter}: Conceptualization; Resources; Supervision; Writing - review \& editing.
\textbf{Chiara Panosetti}: Conceptualization; Methodology; Project administration; Supervision; Writing - review \& editing.


\section*{Conflicts of interest}
There are no conflicts to declare.

\section*{Data and Code Availability}

All code and data will be made available upon publication.

\begin{acknowledgments}

We thank Julian Holland, Artem Samtsevich, Yihua Song, and Christoph Scheurer for helpful discussions.
We jointly and gratefully acknowledge the Max Planck Computing and Data Facility (MPCDF) for providing computational resources and the Bundesministerium für Forschung, Technologie und Raumfahrt (BMFTR) for funding through the ASCEND grant.
\end{acknowledgments}

\bibliographystyle{apsrev4-2}
\bibliography{bibliography}

\end{document}


\begin{center}
{\large\bfseries Supplementary Material: Learning to Converge: Warm-Starting DFTB Self-Consistent Charges with Machine Learning}\\[0.8em]
Maximilian L. Ach, Karsten Reuter, Chiara Panosetti\\[0.3em]
\small Fritz-Haber-Institut der Max-Planck-Gesellschaft, Faradayweg 4-6, D-14195 Berlin, Germany\\
\small panosetti@fhi.mpg.de
\end{center}
\vspace{1em}

\section{DFTB Reference Calculations}
\label{sec:dftb_ref}

All reference DFTB charges are computed using DFTB$+$ \cite{hourahine2020dftb+, hourahine2025recent} (version 24.1).
The SCC procedure is run with a convergence threshold of $10^{-6}$\,e and a maximum of 2,000 SCC cycles, using Broyden charge mixing, the \texttt{RelativelyRobust} eigensolver, and no Fermi smearing (zero electronic temperature).
Unless stated otherwise, all calculations employ the second-order (DFTB2) formulation.

\textbf{QM9, water clusters, solvated amino acids, dipeptides.}
The 3ob-3-1 (hereafter referred to as 3ob) parameter set\cite{gaus2013parametrization, gaus2014parameterization, kubillus2015parameterization} is used throughout.
For the SPICE subsets, structures with a non-zero total charge are discarded prior to reference calculations, as only neutral systems are targeted in this work.

\textbf{LLZO.}
Structures are 192-atom cubic supercells of the garnet unit cell.
A $2\times2\times2$ Monkhorst-Pack $k$-point grid is used.
The PTBP parameter set\cite{cui2024obtaining} is used.

\textbf{Ni$_x$O$_y$.}
Structures are periodic and contain up to 16 atoms.
A fixed $3\times3\times3$ Monkhorst-Pack $k$-point grid is used for all structures.
The mio-1-1 (hereafter referred to as mio) parameter set\cite{elstner1998self} covers O, and the trans3d-0-1 extension\cite{zheng2007parameter} covers Ni and Ni--O interactions.
All calculations are non-spin-polarized.

\textbf{Convergence failures.}
A small number of reference calculations did not converge within the 2,000-cycle limit and were excluded from the training data.
For QM9, 3 structures failed to converge.
For the dipeptides subset, 23 structures failed to converge.
For Ni$_x$O$_y$, 1,118 structures failed to converge.
All calculations for the water clusters, solvated amino acids, and LLZO datasets converged.
The effect of ML-predicted charge initialization on these previously failing structures is reported in the main text.

\section{Model Training Details}
\label{sec:model_training}

\subsection{SOAP Descriptor Hyperparameters}

SOAP descriptors are computed using the DScribe library (version 2.1.2).
The local atomic environment of each atom is described up to a cutoff radius $r_\text{cut}$, with atomic positions smeared by Gaussians of width $\sigma_\text{atom}$.
The neighbor density is expanded in $n_\text{max}$ radial basis functions and spherical harmonics up to $l_\text{max}$.
The parameters are kept fixed across all datasets and are summarized in Table~\ref{tab:soap_params}.

\begin{table}[H]
    \centering
    \caption{SOAP descriptor hyperparameters used for all datasets.}
    \label{tab:soap_params}
    \begin{tabular}{lr}
        \toprule
        Parameter & Value \\
        \midrule
        $r_\text{cut}$ (\AA) & 6.0 \\
        $\sigma_\text{atom}$ (\AA) & 1.0 \\
        $n_\text{max}$ & 12 \\
        $l_\text{max}$ & 10 \\
        \bottomrule
    \end{tabular}
\end{table}

\newpage
\subsection{KRR Hyperparameter Selection}

For each element-specific model, 20\% of the data is held out as a test set and not used during hyperparameter selection.
The KRR regularization parameter $\lambda$ and the RBF kernel bandwidth $\sigma$ are then selected by five-fold cross-validation grid search over a logarithmic grid on the remaining data.
The search ranges are given in Table~\ref{tab:krr_grid}.

\begin{table}[H]
\centering
\caption{Hyperparameter grid used for KRR model selection, grouped by dataset.}
\label{tab:krr_grid}
\begin{tabular}{llll}
\toprule
Dataset(s) & Parameter & Range & Values \\
\midrule
\multirow{2}{*}{QM9, water clusters, solvated amino acids, dipeptides}
 & $\lambda$ & $[10^{-8},\, 10^{-6}]$ & 3 (log-spaced) \\
 & $\sigma$  & $[10^{-8},\, 10^{-5}]$ & 4 (log-spaced) \\
\midrule
\multirow{2}{*}{LLZO, Ni$_x$O$_y$}
 & $\lambda$ & $[10^{-8},\, 10^{-5}]$ & 4 (log-spaced) \\
 & $\sigma$  & $[10^{-9},\, 10^{-5}]$ & 5 (log-spaced) \\
\bottomrule
\end{tabular}
\end{table}

\section{Charge Prediction Accuracy}

\subsection{DFTB Models}

\begin{figure}[H]
\centering
\includegraphics[width=\textwidth]{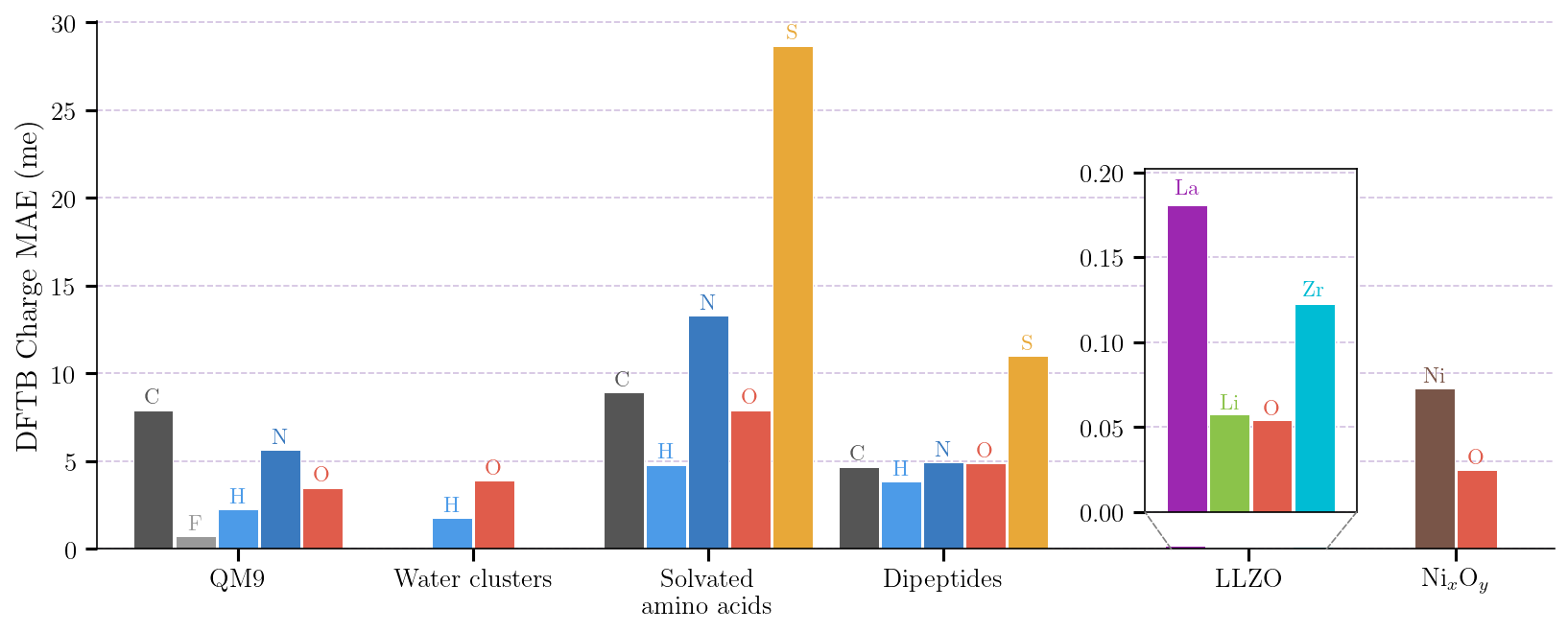}
\caption{DFTB charge prediction MAE for all elements across datasets. The LLZO inset uses millielectron ($me$) units. Compare with the RMSE results in the main text.}
\label{fig:mae_barchart}
\end{figure}

\begin{figure}[H]
\centering
\includegraphics[width=\textwidth]{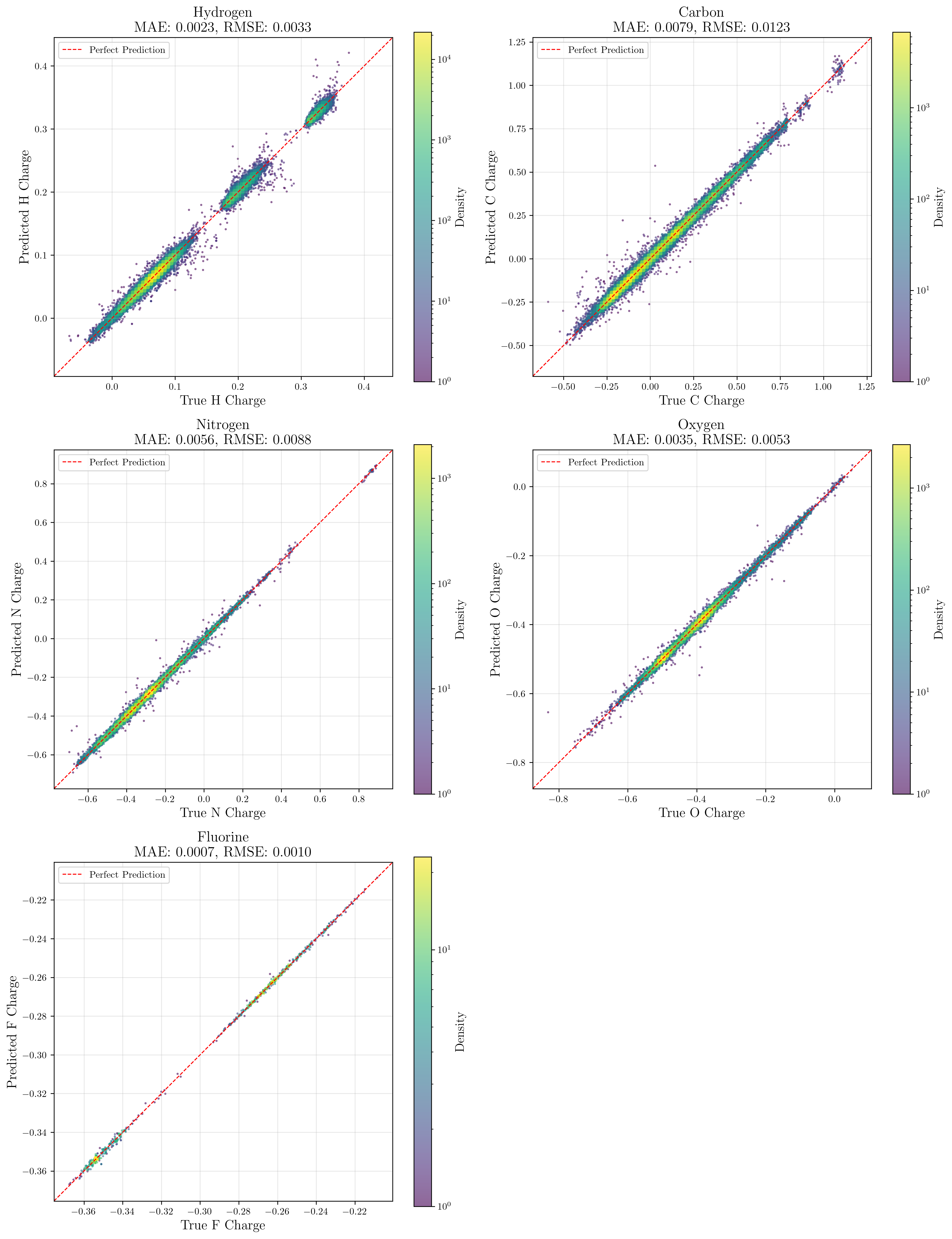}
\caption{Parity plots of ML-predicted vs.\ reference DFTB charges for QM9 (H, C, N, O, F). Color indicates point density.}
\label{fig:parity_qm9}
\end{figure}

\begin{figure}[H]
\centering
\includegraphics[width=\textwidth]{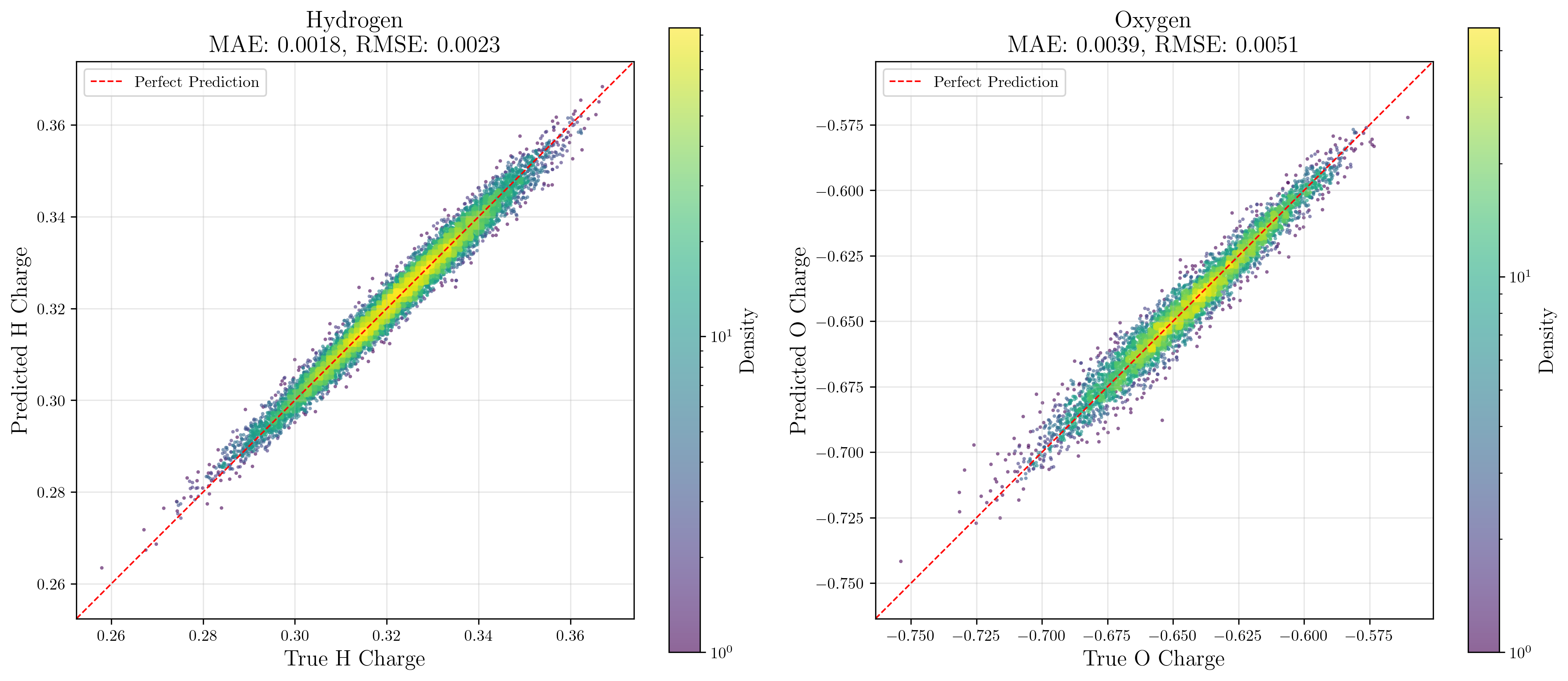}
\caption{Parity plots of ML-predicted vs.\ reference DFTB charges for water clusters (H, O).}
\label{fig:parity_water}
\end{figure}

\begin{figure}[H]
\centering
\includegraphics[width=\textwidth]{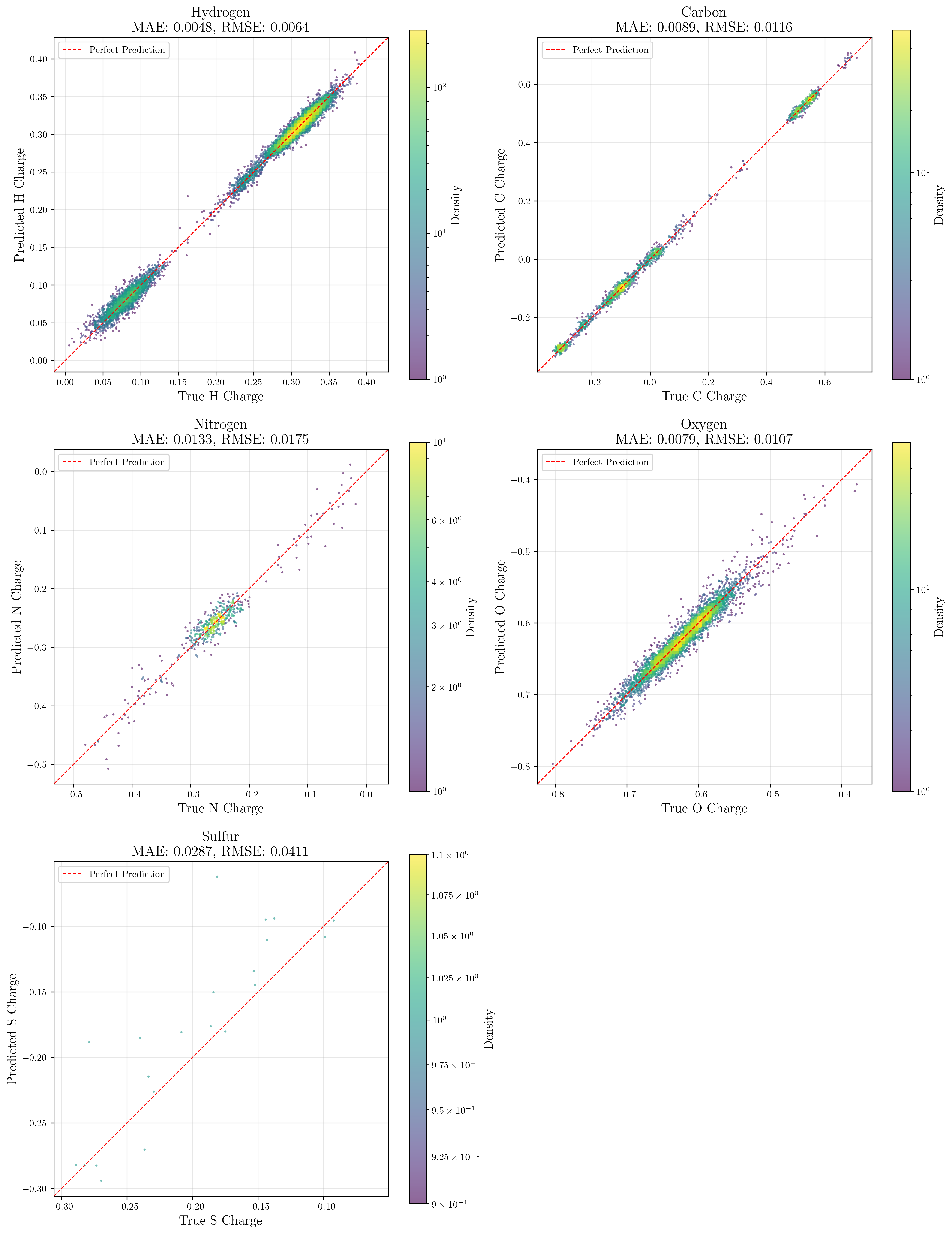}
\caption{Parity plots of ML-predicted vs.\ reference DFTB charges for solvated amino acids (H, C, N, O, S).}
\label{fig:parity_solaa}
\end{figure}

\begin{figure}[H]
\centering
\includegraphics[width=\textwidth]{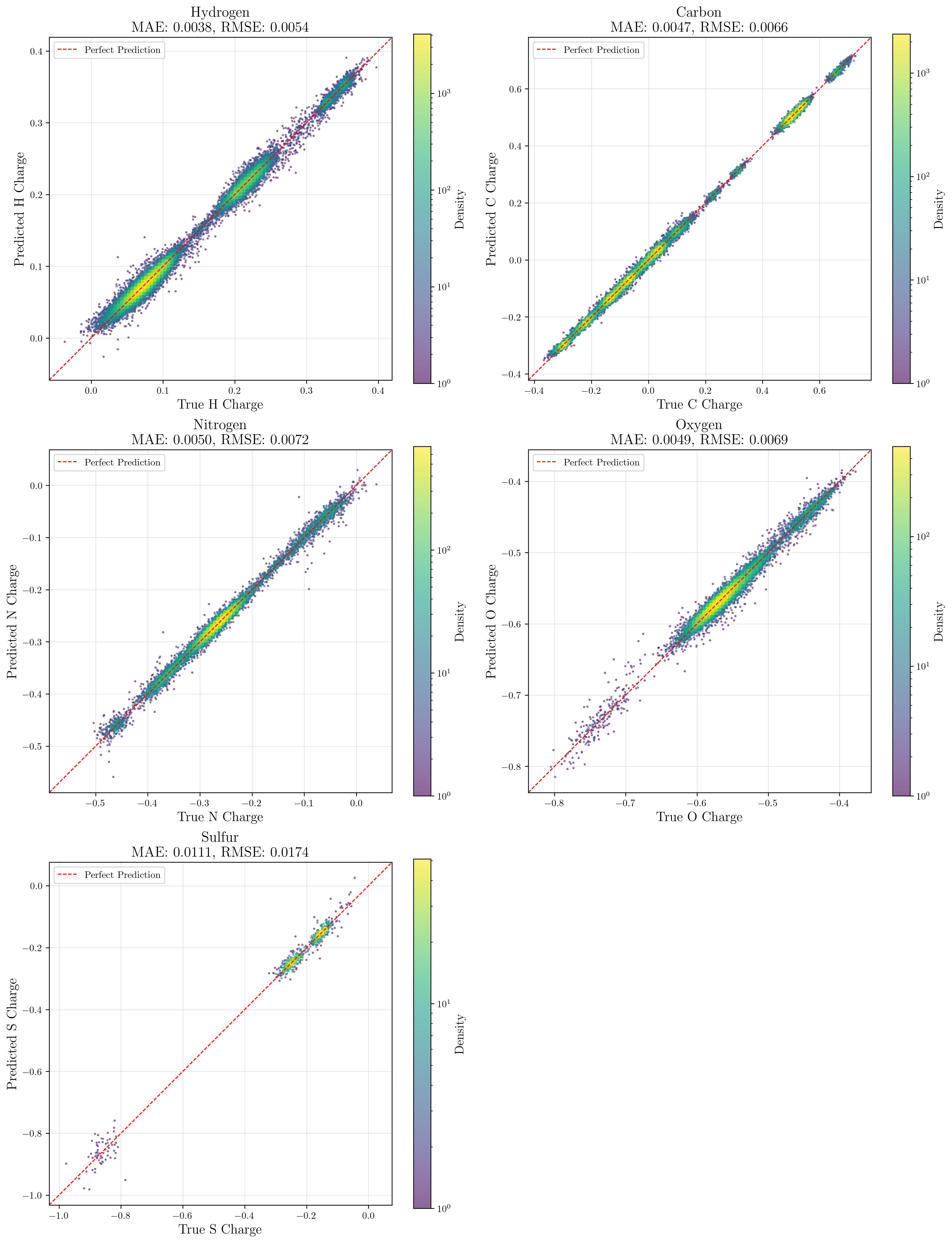}
\caption{Parity plots of ML-predicted vs.\ reference DFTB charges for dipeptides (H, C, N, O, S).}
\label{fig:parity_dipep}
\end{figure}

\begin{figure}[H]
    \centering
    \includegraphics[width=\textwidth]{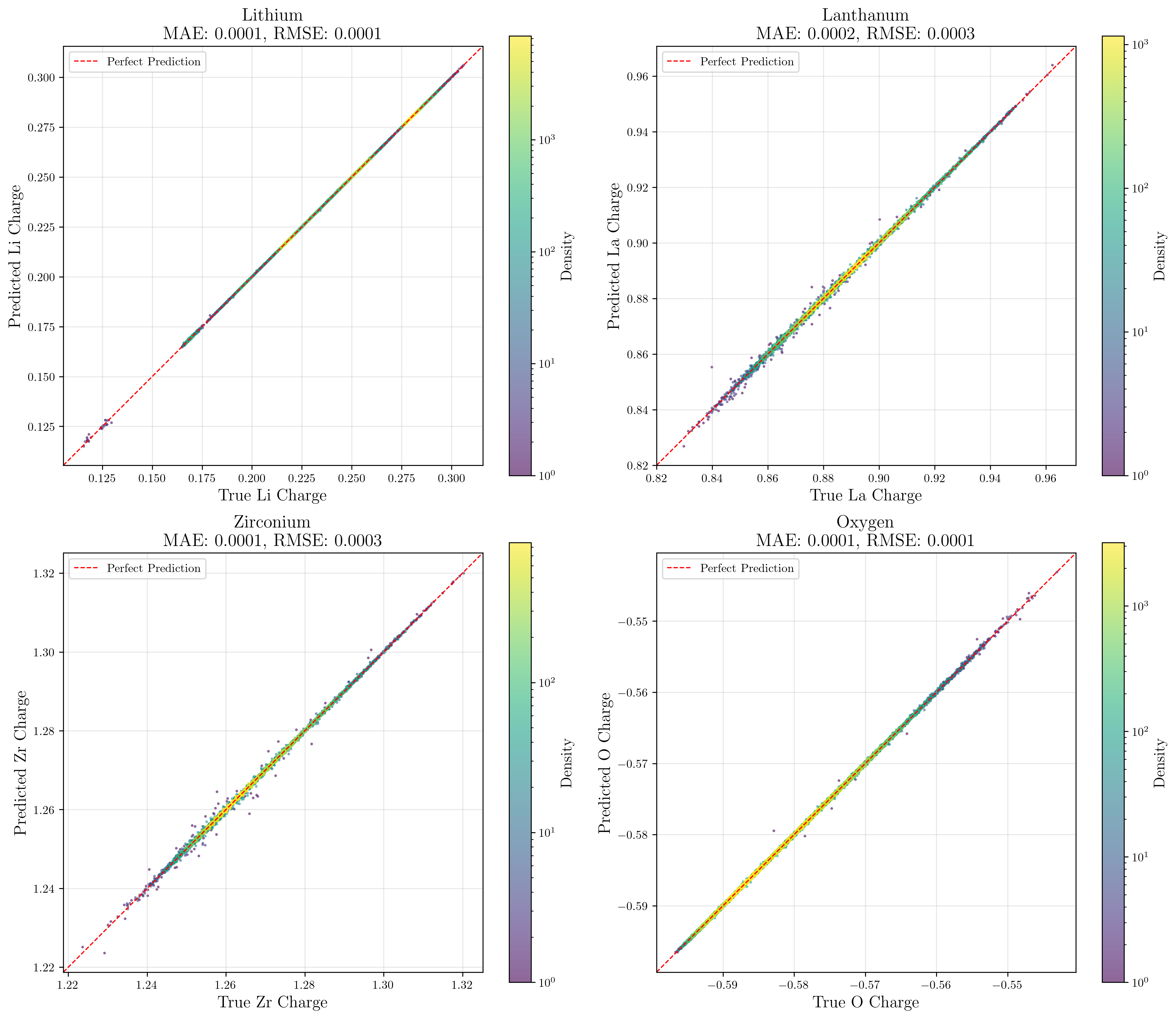}
    \caption{Parity plots of ML-predicted vs.\ reference DFTB charges for LLZO (Li, La, Zr, O).}
    \label{fig:parity_llzo}
\end{figure}

\begin{figure}[H]
\centering
\includegraphics[width=\textwidth]{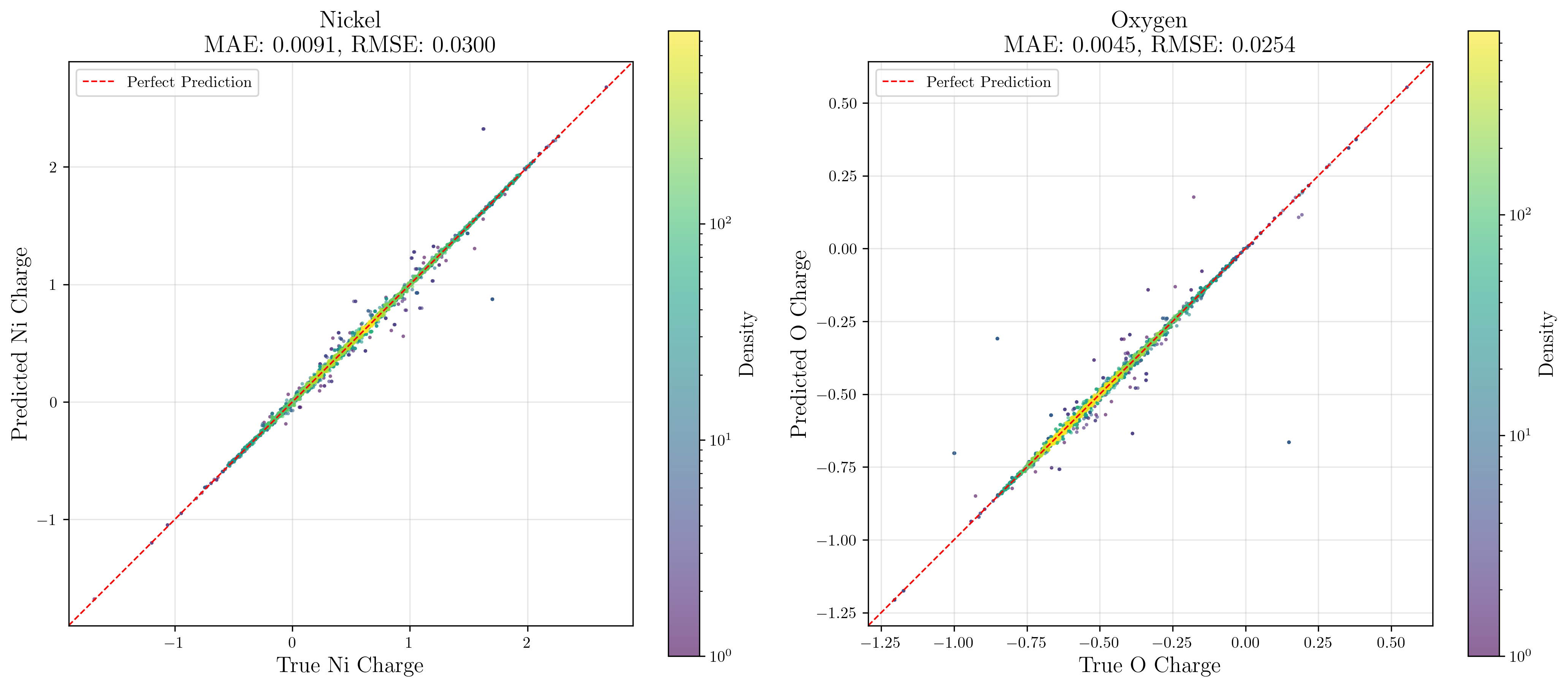}
\caption{Parity plots of ML-predicted vs.\ reference DFTB charges for Ni$_x$O$_y$.}
\label{fig:parity_nio}
\end{figure}

\subsection{DFT Models}

\begin{figure}[H]
\centering
\includegraphics[width=\textwidth]{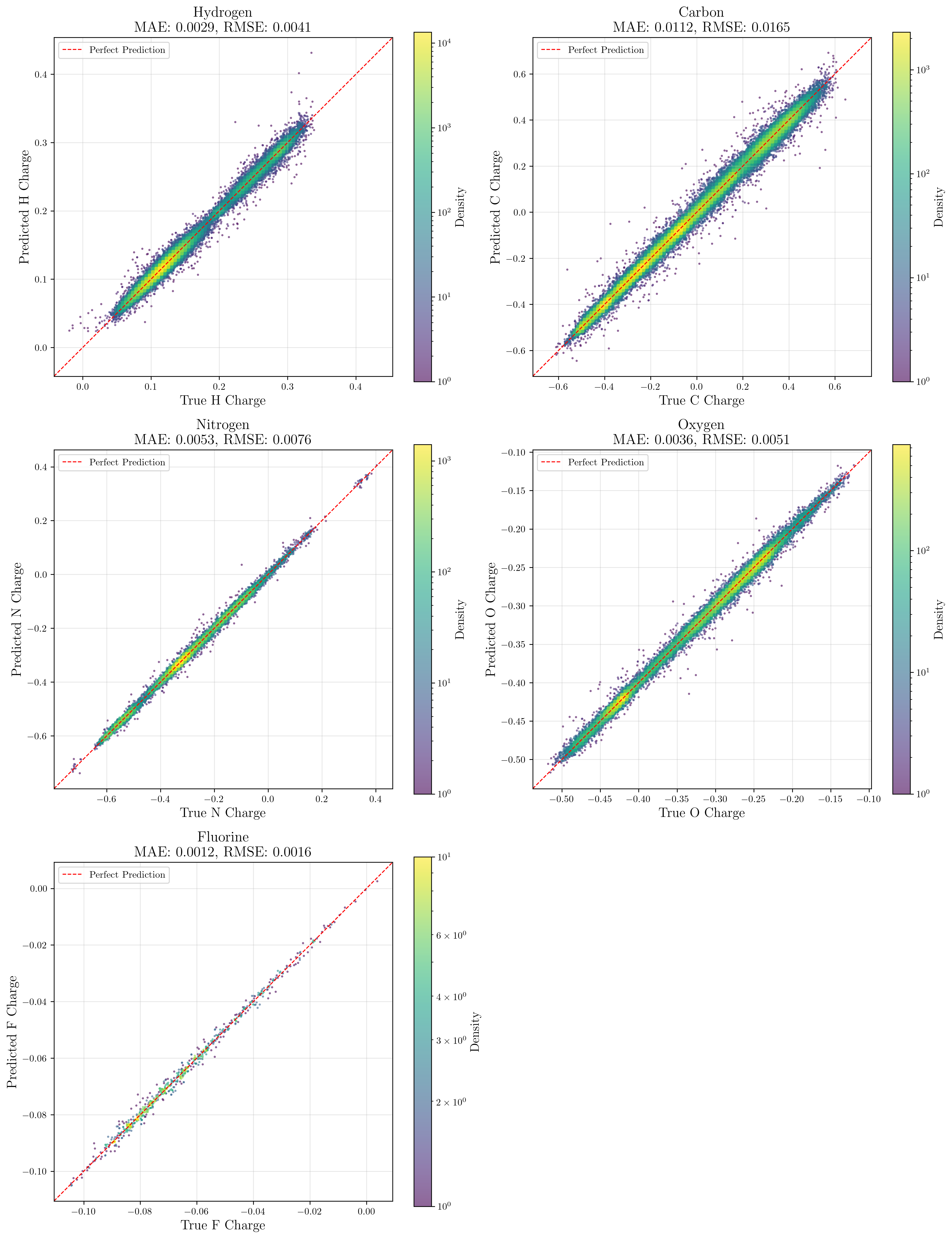}
\caption{Parity plots of ML-predicted vs.\ reference DFT charges for QM9 (H, C, N, O, F). Color indicates point density.}
\label{fig:parity_qm9_dft}
\end{figure}

\begin{figure}[H]
\centering
\includegraphics[width=\textwidth]{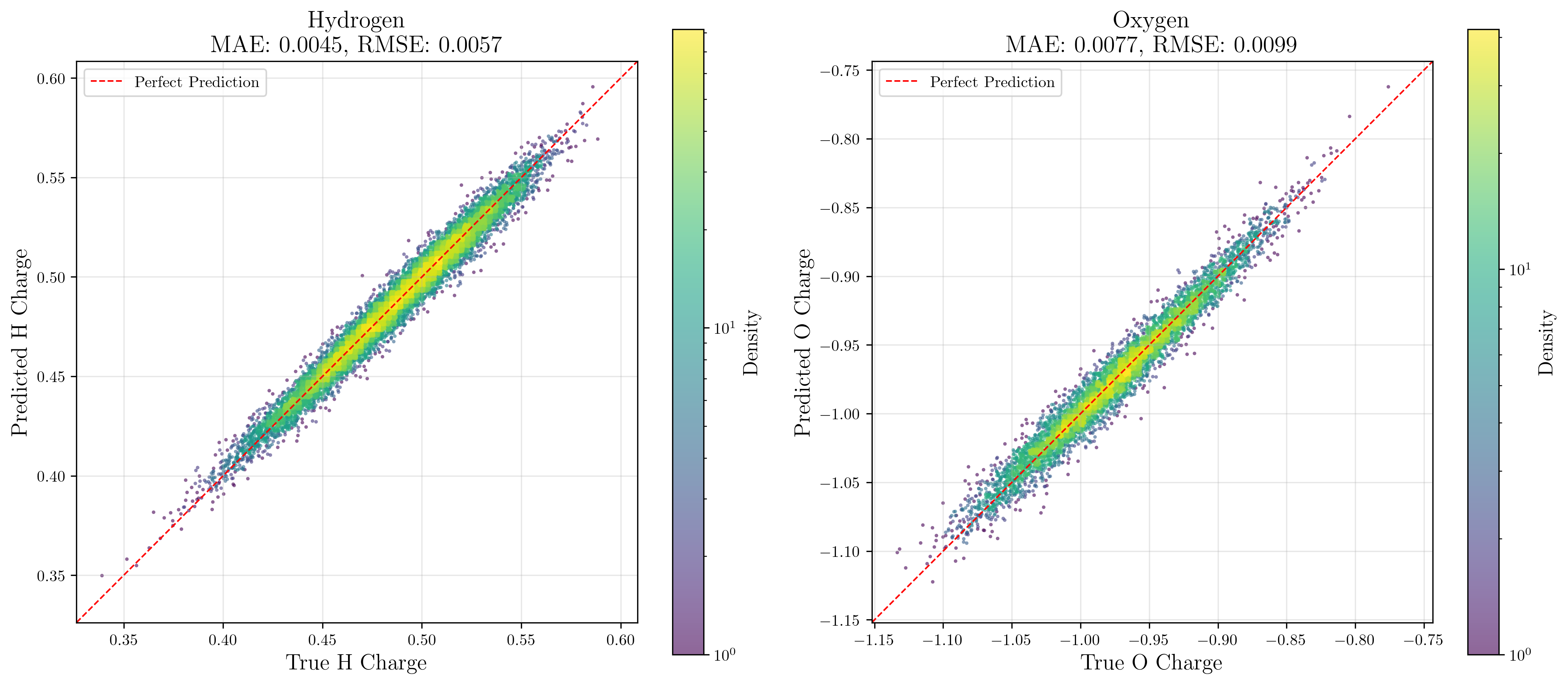}
\caption{Parity plots of ML-predicted vs.\ reference DFT charges for water clusters (H, O).}
\label{fig:parity_water_dft}
\end{figure}

\begin{figure}[H]
\centering
\includegraphics[width=\textwidth]{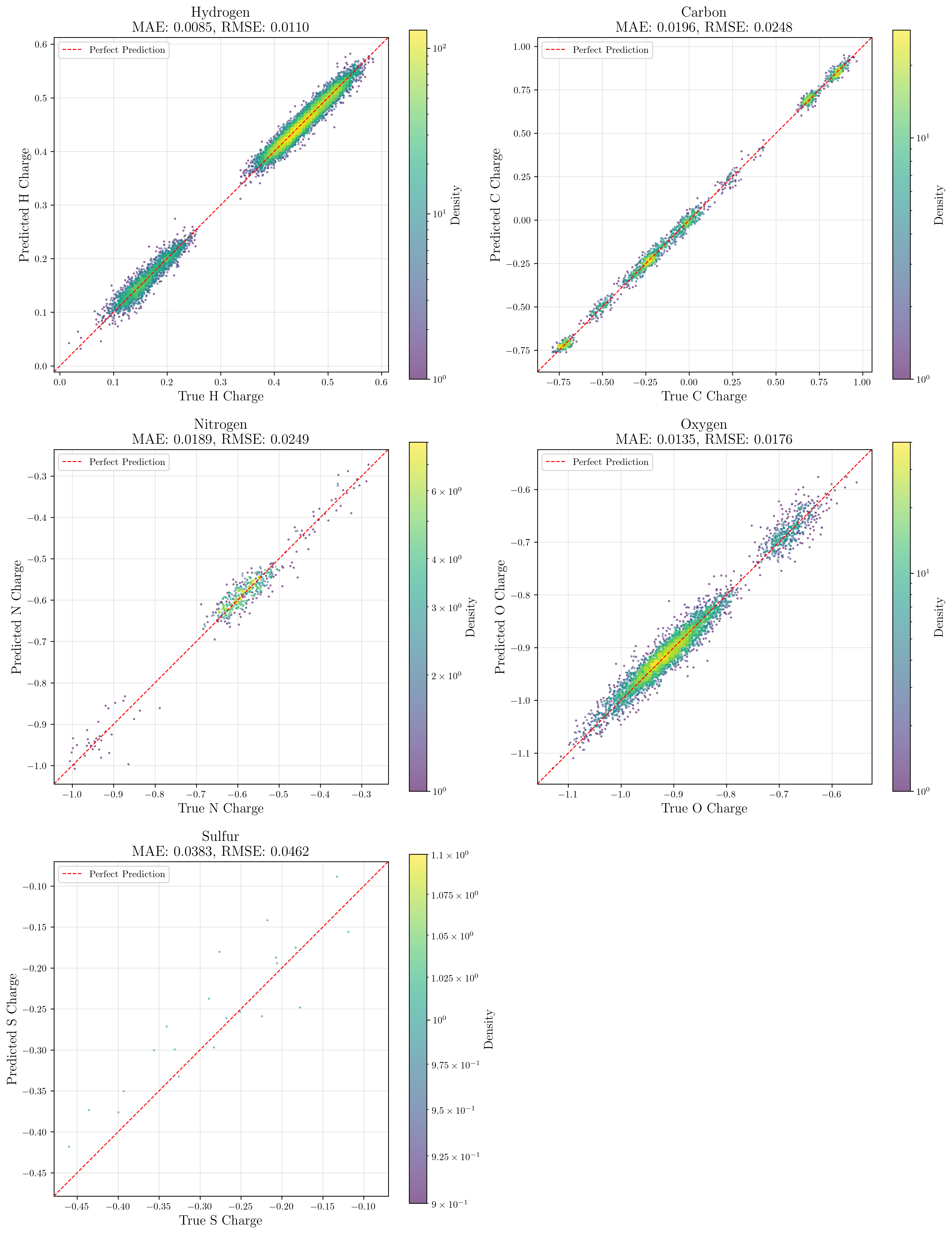}
\caption{Parity plots of ML-predicted vs.\ reference DFT charges for solvated amino acids (H, C, N, O, S).}
\label{fig:parity_solaa_dft}
\end{figure}

\begin{figure}[H]
\centering
\includegraphics[width=\textwidth]{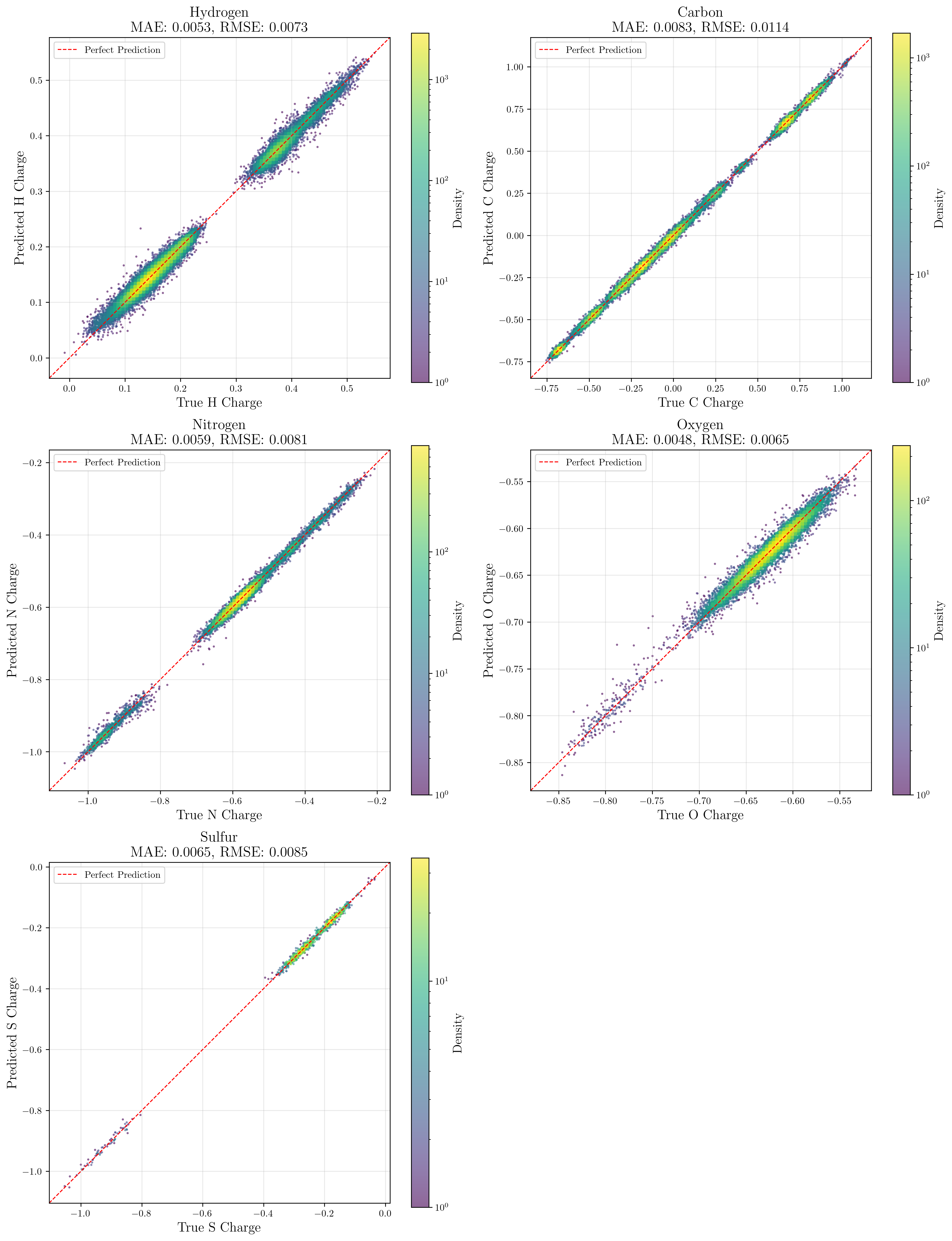}
\caption{Parity plots of ML-predicted vs.\ reference DFT charges for dipeptides (H, C, N, O, S).}
\label{fig:parity_dipep_dft}
\end{figure}

\begin{figure}[H]
    \centering
    \includegraphics[width=\textwidth]{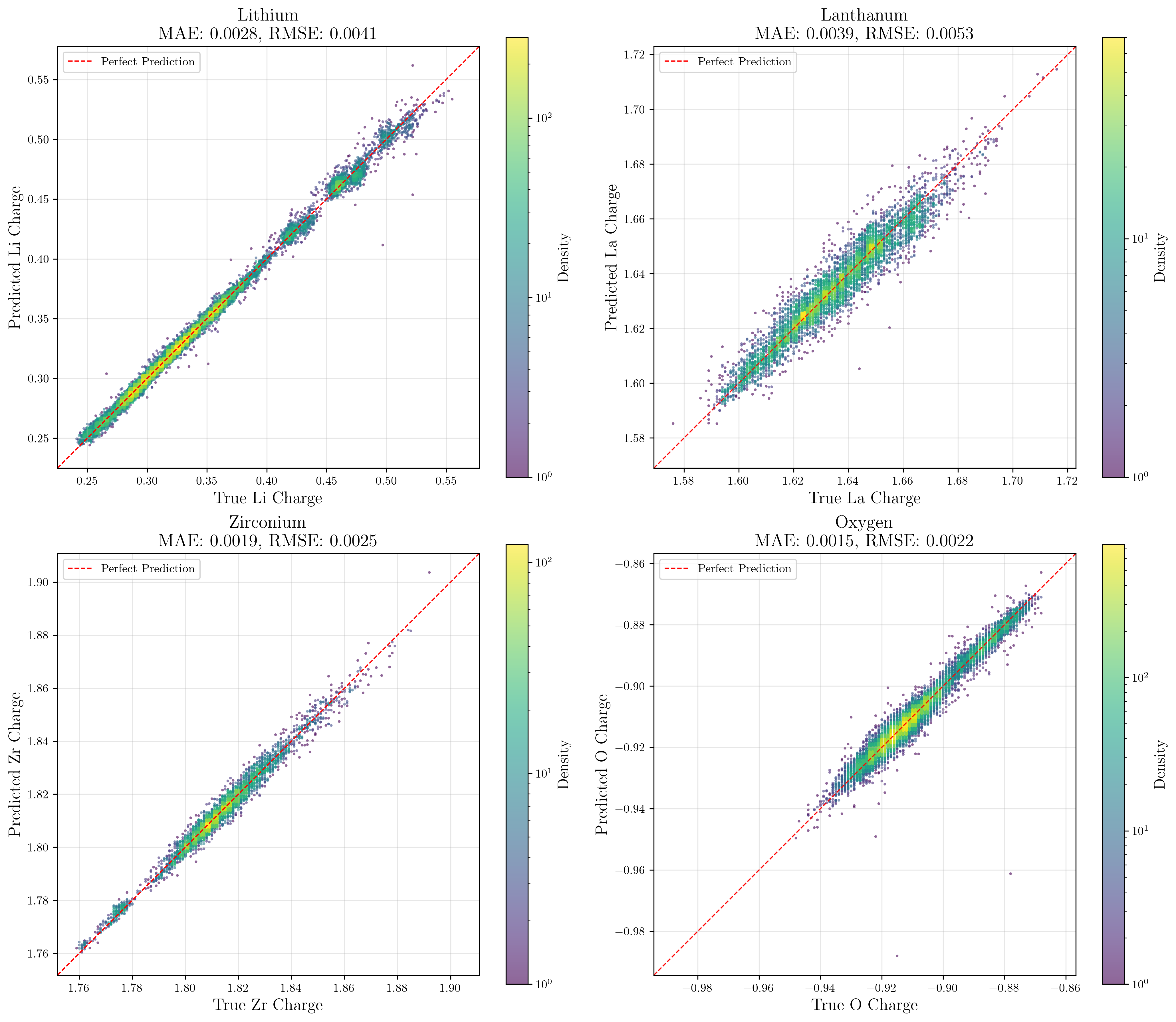}
    \caption{Parity plots of ML-predicted vs.\ reference DFT charges for LLZO (Li, La, Zr, O). The vertical stripe pattern visible for La, O, and Zr arises from the limited three-decimal-place precision of the Mulliken charges in the ONETEP output, which discretises the true-charge axis at 0.001\,e intervals. Li is less affected because its charges span a wider range (${\sim}0.31$\,e) owing to the two distinct coordination environments (tetrahedral and octahedral sites) present in the garnet structure.}
    \label{fig:parity_llzo_dft}
\end{figure}

\begin{figure}[H]
\centering
\includegraphics[width=\textwidth]{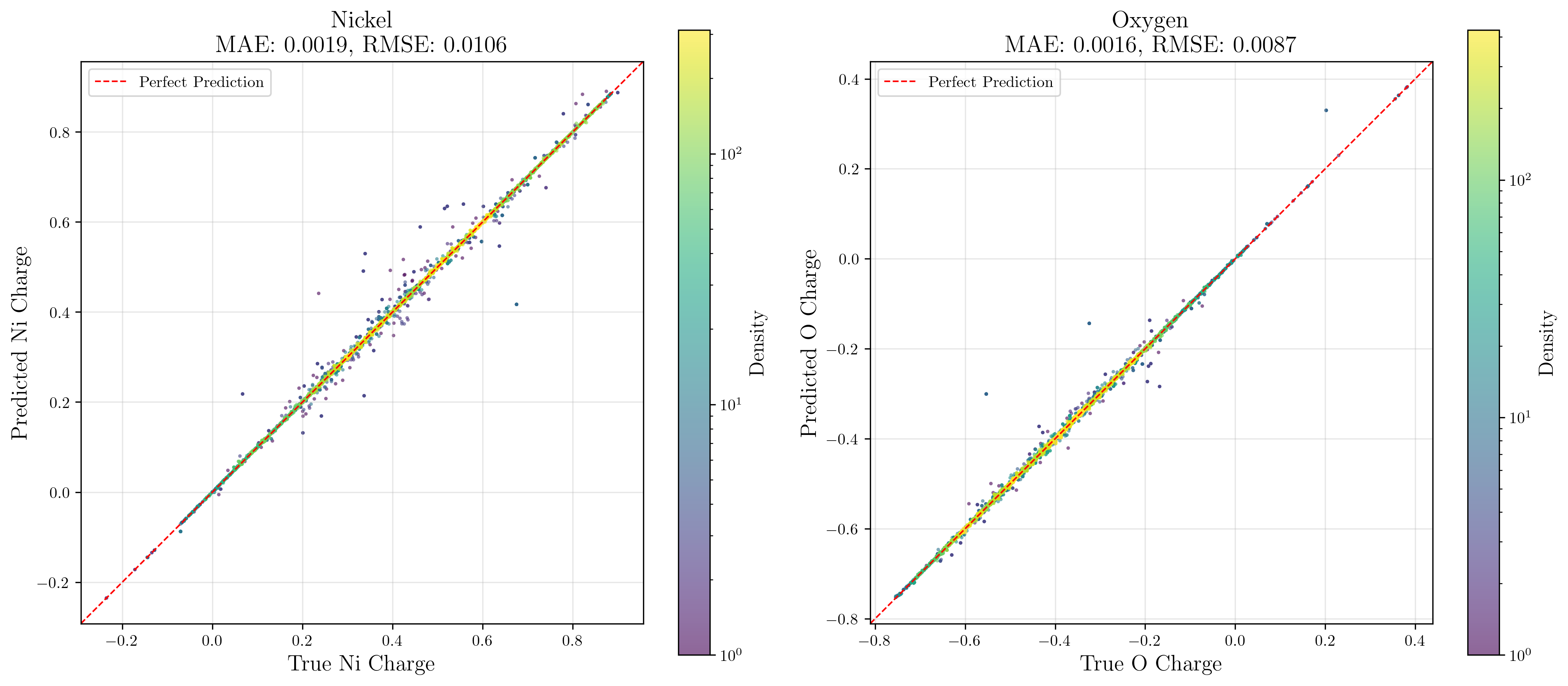}
\caption{Parity plots of ML-predicted vs.\ reference DFT charges for Ni$_x$O$_y$.}
\label{fig:parity_nio_dft}
\end{figure}

\section{Learning Curves}
\label{sec:learning_curves}

\subsection{DFTB Models}

\begin{figure}[H]
\centering
\includegraphics[width=\textwidth]{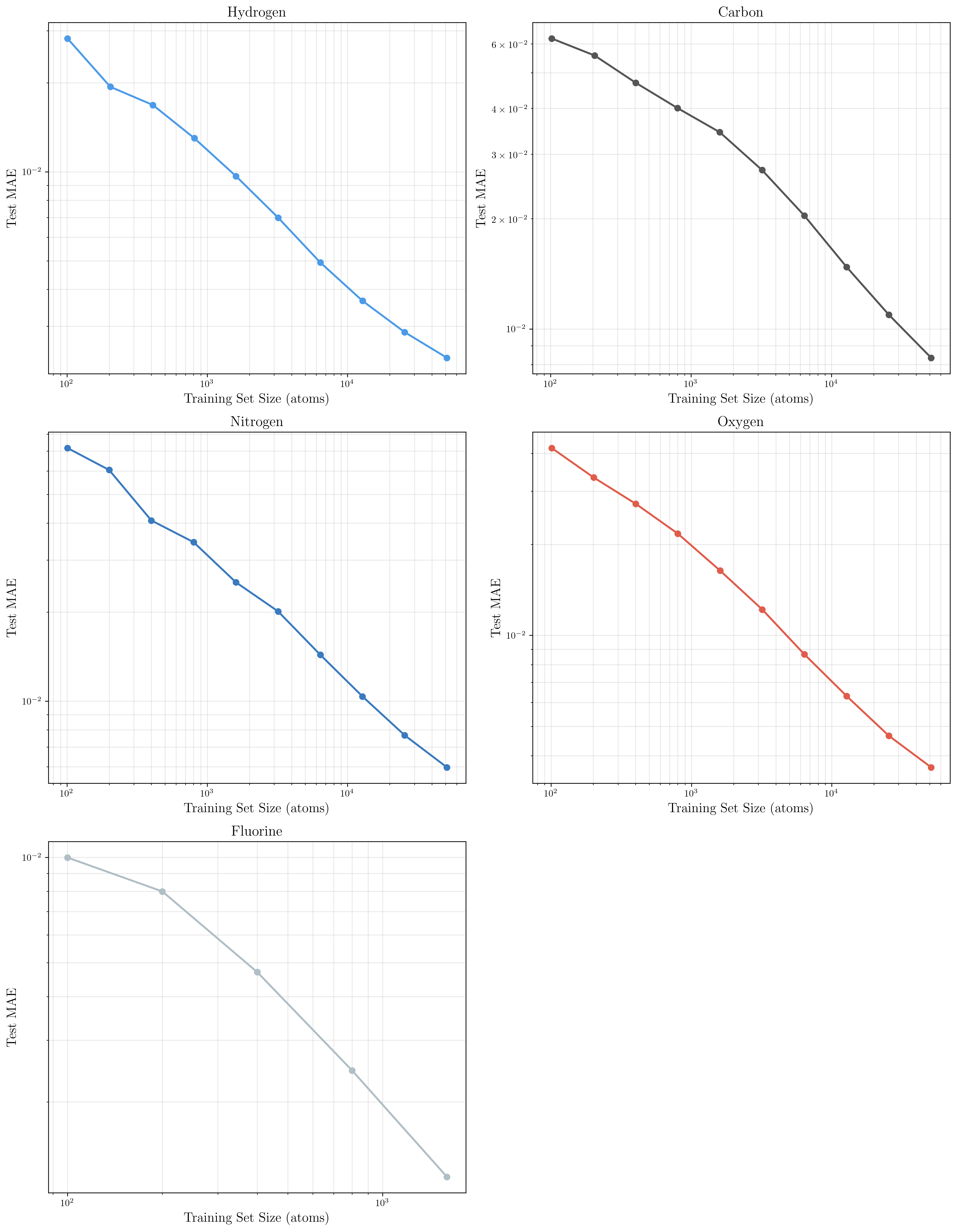}
\caption{Learning curves for QM9 (H, C, N, O, F): validation MAE as a function of training set size.}
\label{fig:lc_qm9}
\end{figure}

\begin{figure}[H]
\centering
\includegraphics[width=\textwidth]{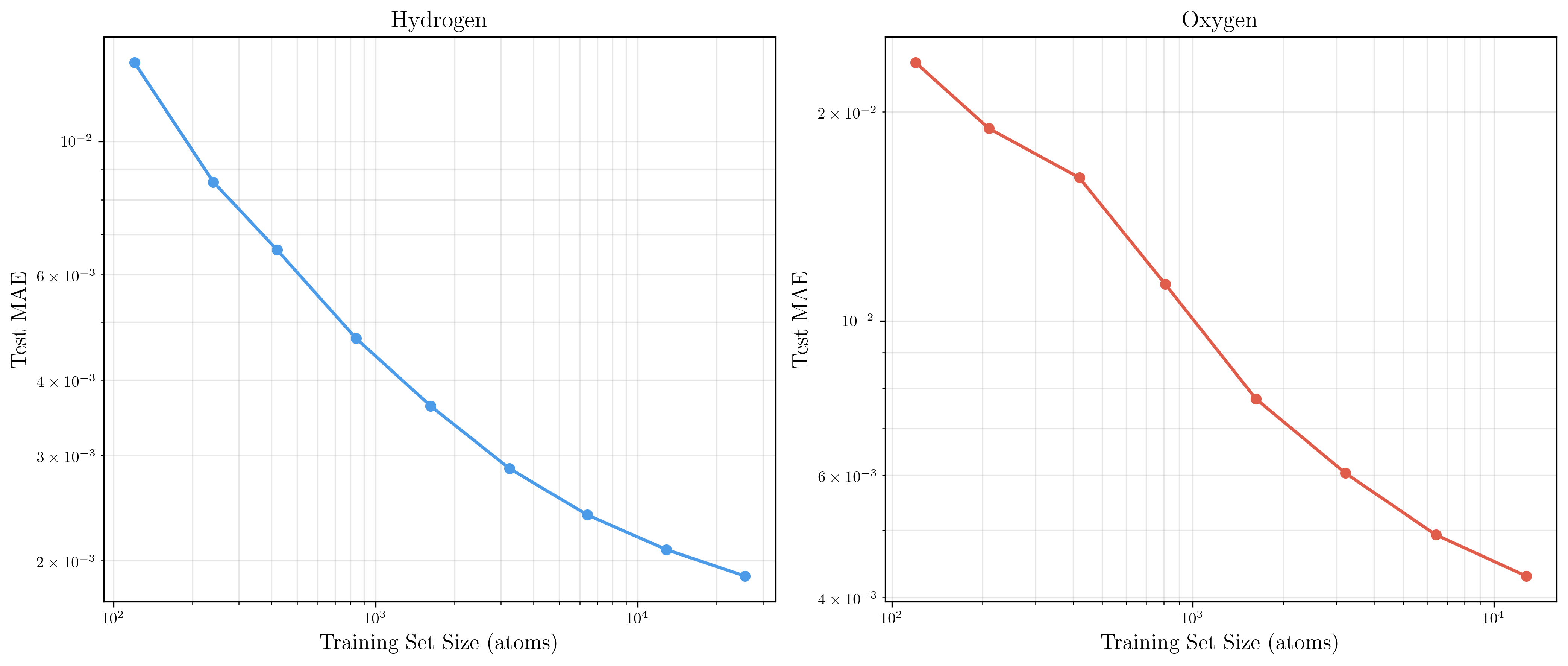}
\caption{Learning curves for water clusters (H, O).}
\label{fig:lc_water}
\end{figure}

\begin{figure}[H]
\centering
\includegraphics[width=\textwidth]{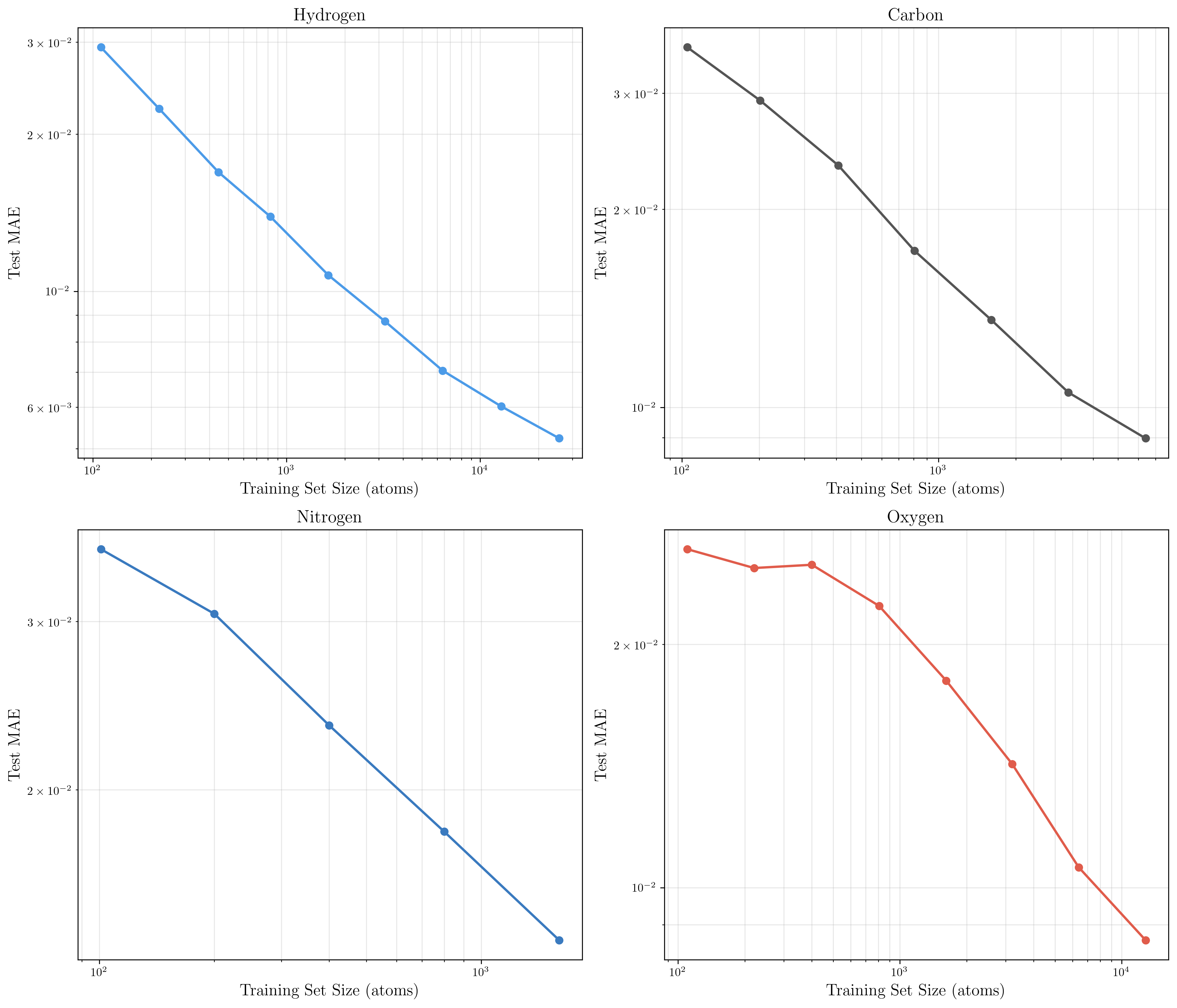}
\caption{Learning curves for solvated amino acids (H, C, N, O). Sulfur is excluded due to insufficient training data.}
\label{fig:lc_solaa}
\end{figure}

\begin{figure}[H]
\centering
\includegraphics[width=\textwidth]{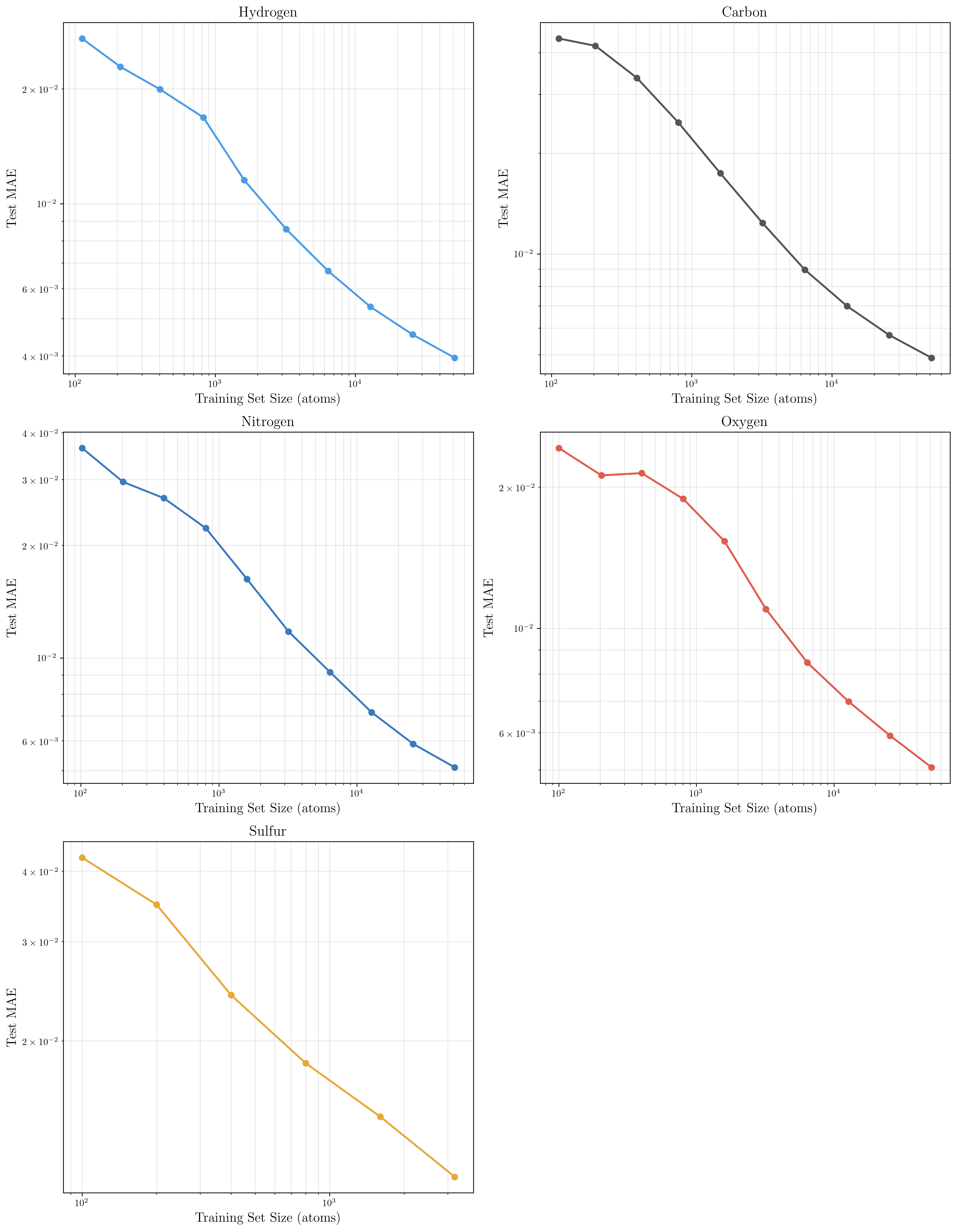}
\caption{Learning curves for dipeptides (H, C, N, O, S).}
\label{fig:lc_dipep}
\end{figure}

\begin{figure}[H]
    \centering
    \includegraphics[width=\textwidth]{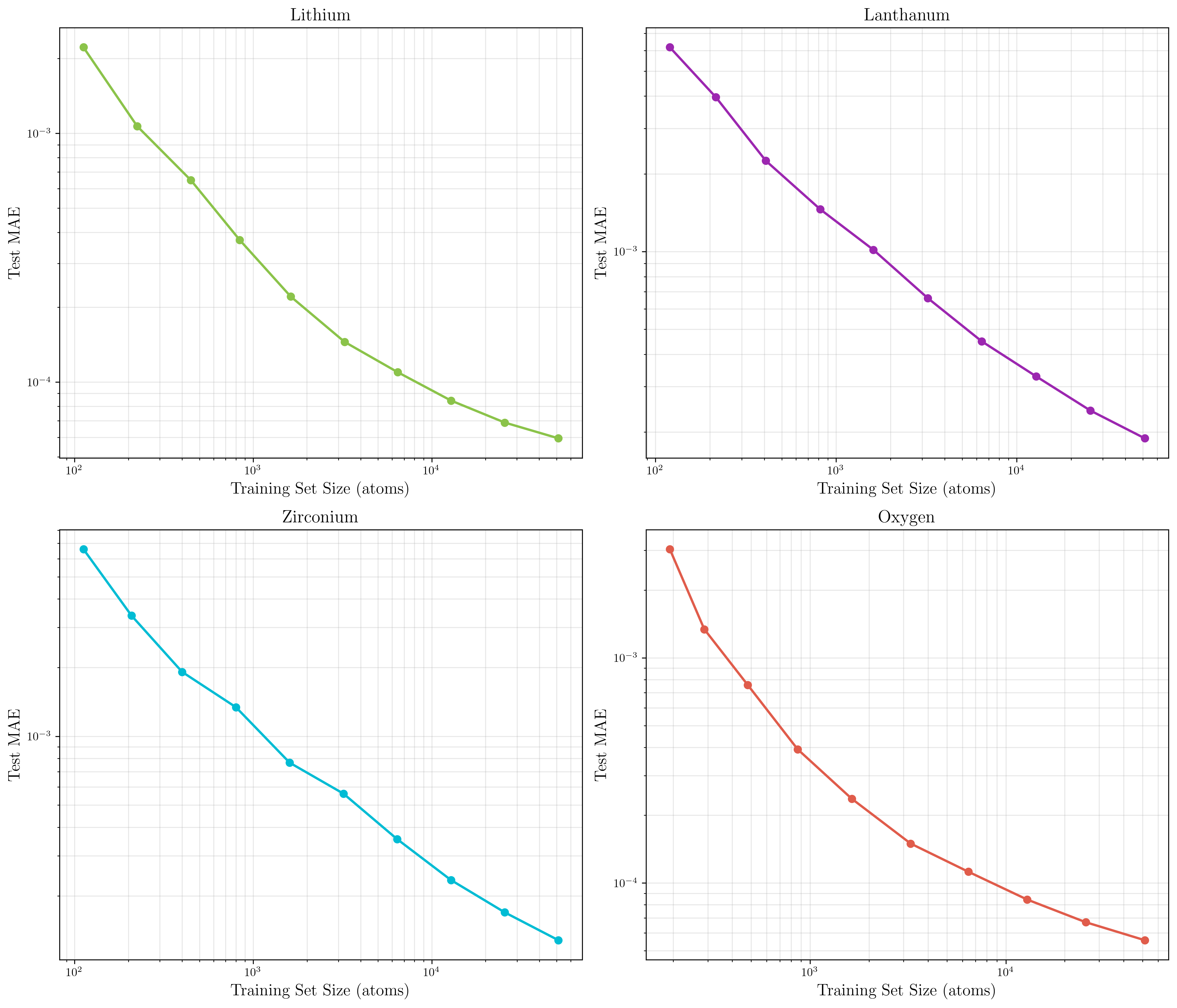}
    \caption{Learning curves for LLZO (Li, La, Zr, O).}
    \label{fig:lc_llzo}
\end{figure}

\begin{figure}[H]
\centering
\includegraphics[width=\textwidth]{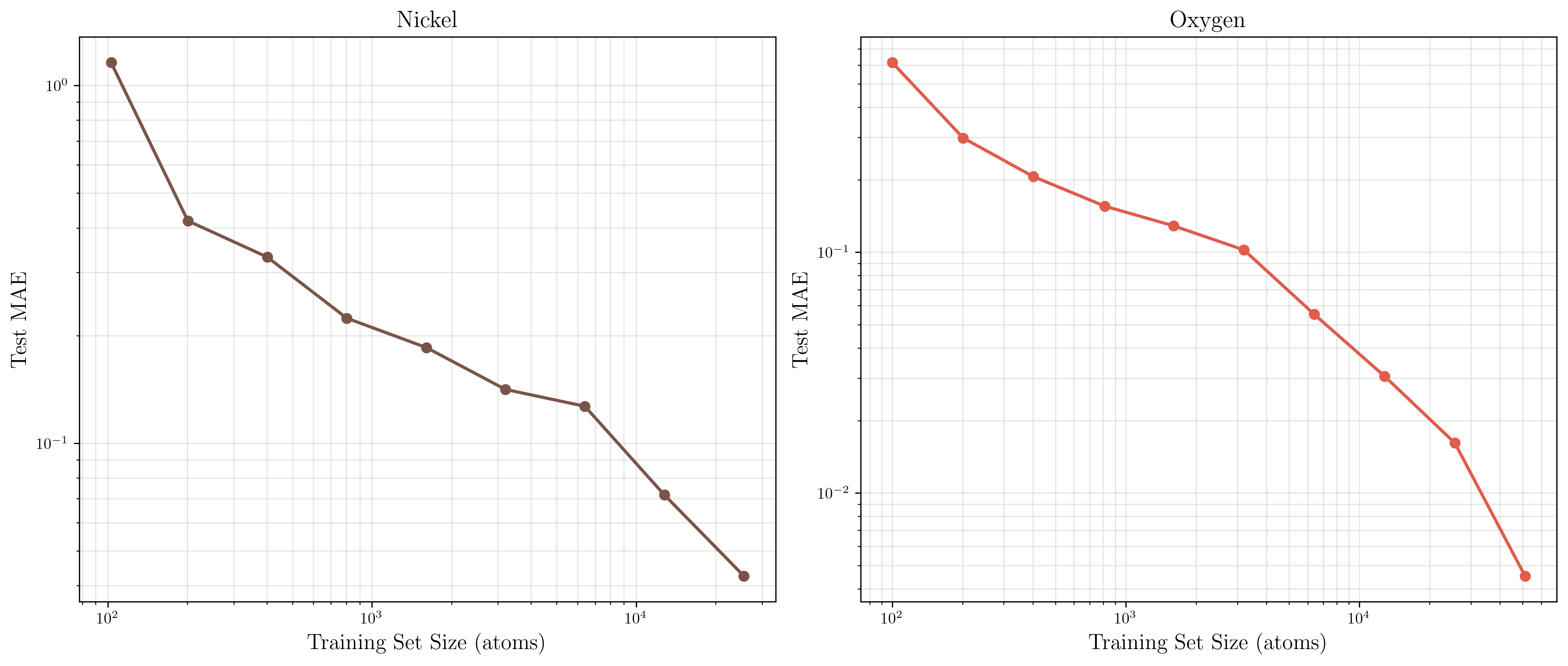}
\caption{Learning curves for Ni$_x$O$_y$.}
\label{fig:lc_nio}
\end{figure}

\subsection{DFT Models}

\begin{figure}[H]
\centering
\includegraphics[width=\textwidth]{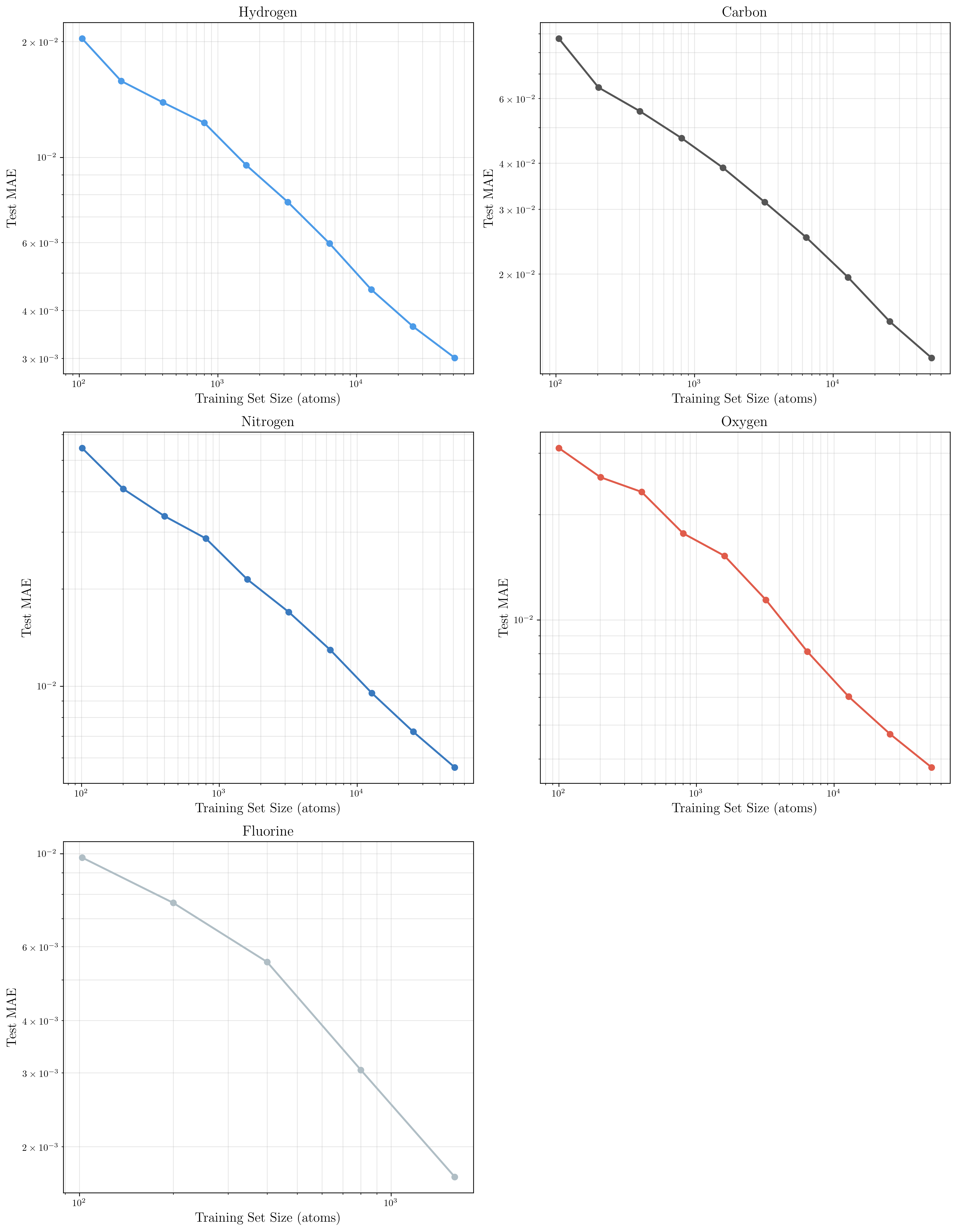}
\caption{Learning curves for QM9 DFT models (H, C, N, O, F): validation MAE as a function of training set size.}
\label{fig:lc_qm9_dft}
\end{figure}

\begin{figure}[H]
\centering
\includegraphics[width=\textwidth]{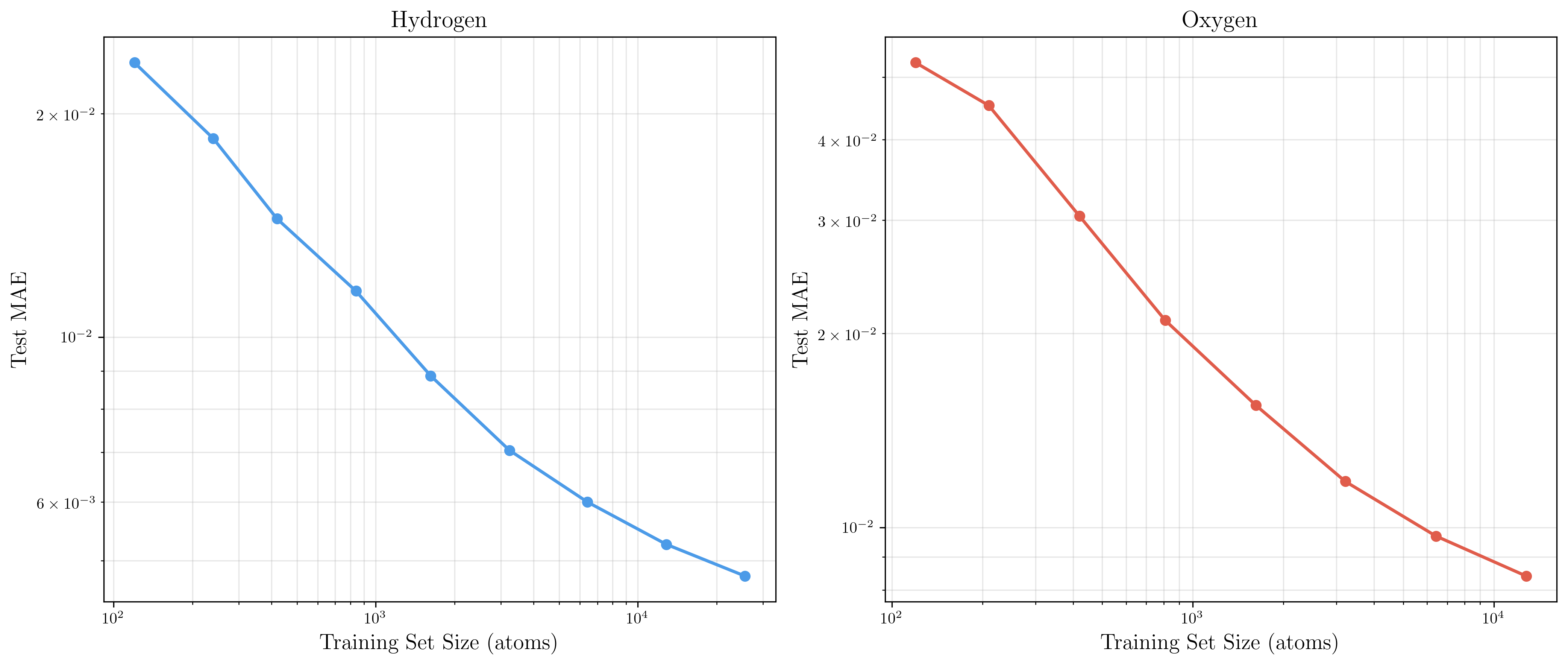}
\caption{Learning curves for water clusters DFT models (H, O).}
\label{fig:lc_water_dft}
\end{figure}

\begin{figure}[H]
\centering
\includegraphics[width=\textwidth]{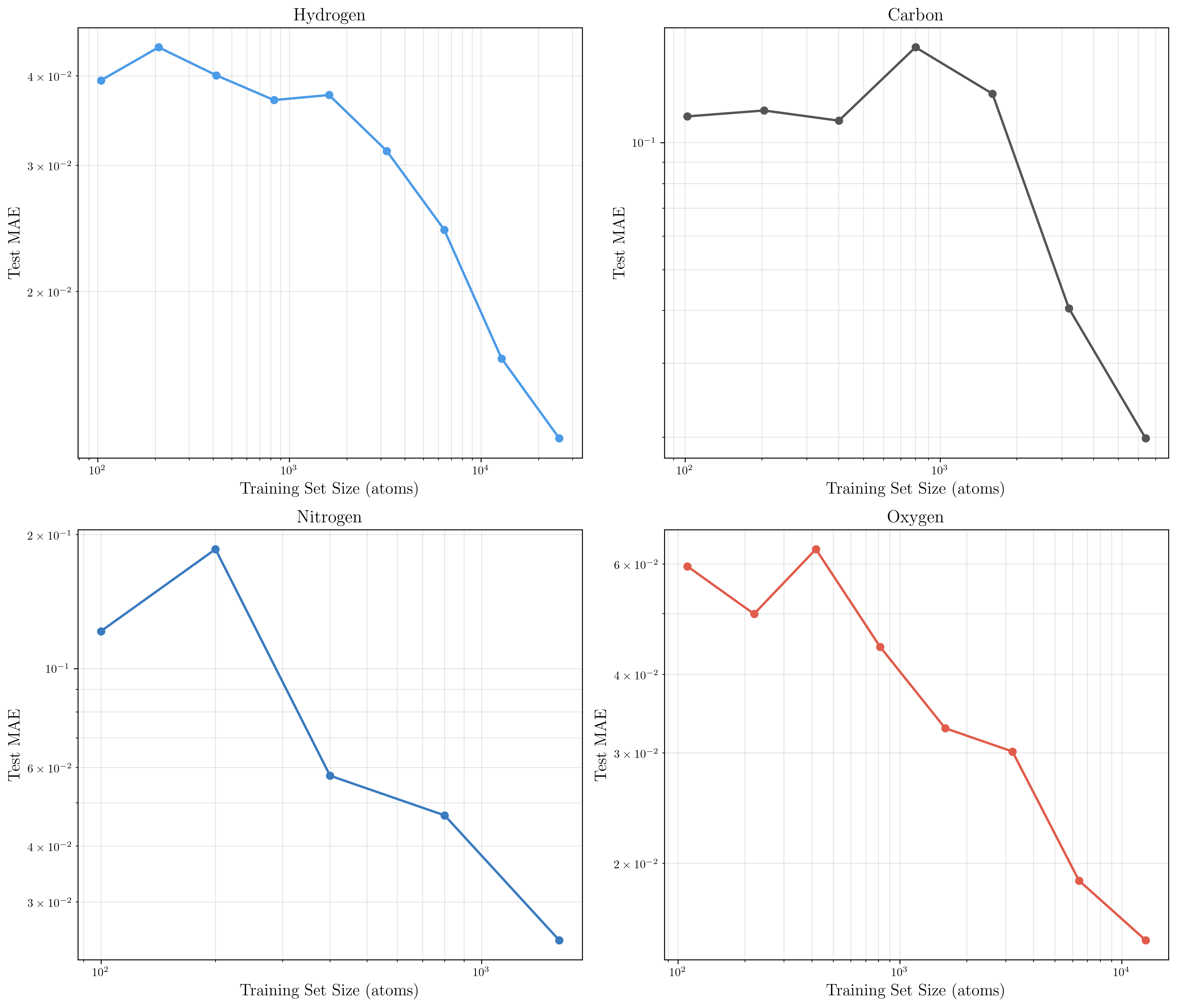}
\caption{Learning curves for solvated amino acids DFT models (H, C, N, O). Sulfur is excluded due to insufficient training data.}
\label{fig:lc_solaa_dft}
\end{figure}

\begin{figure}[H]
\centering
\includegraphics[width=\textwidth]{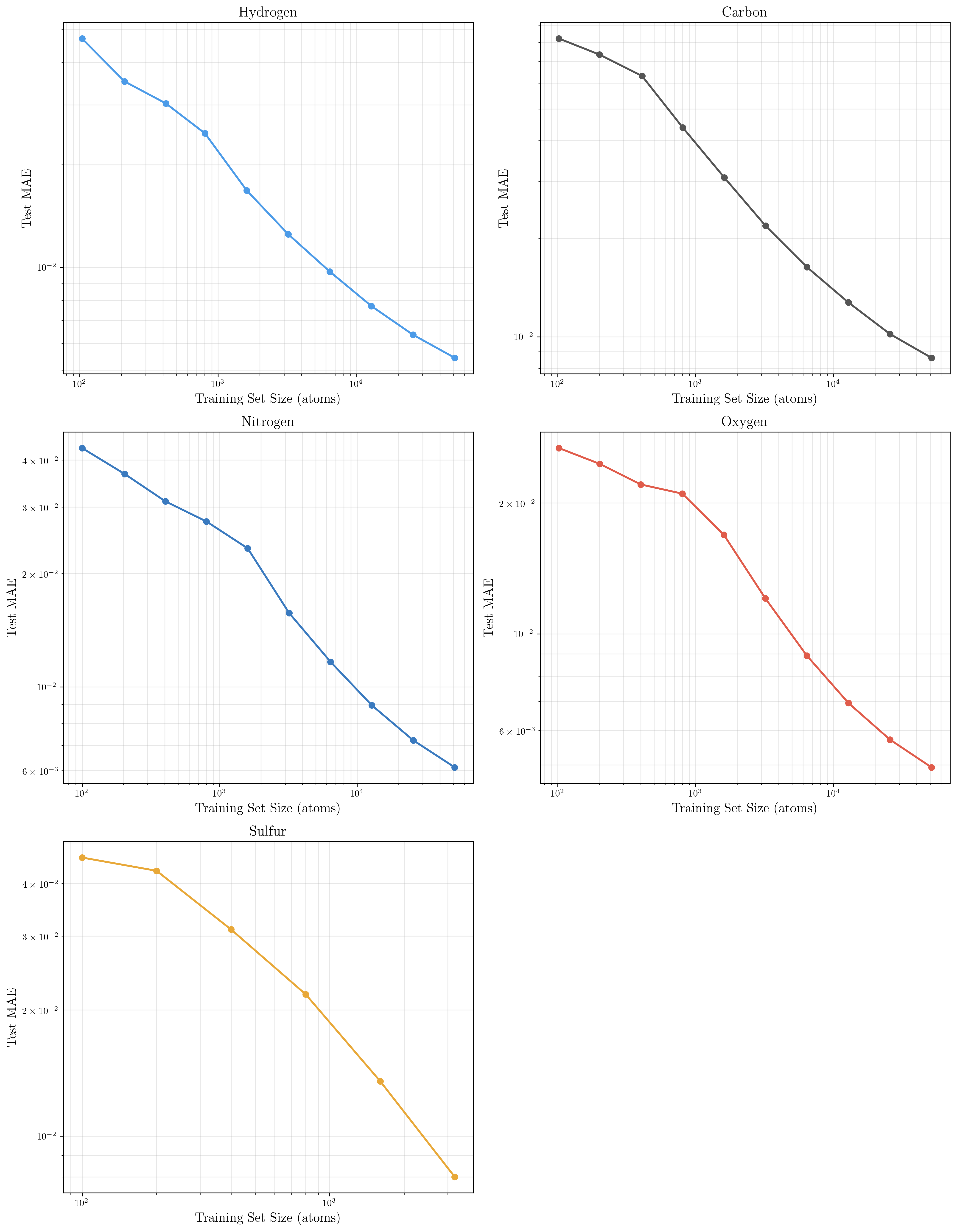}
\caption{Learning curves for dipeptides DFT models (H, C, N, O, S).}
\label{fig:lc_dipep_dft}
\end{figure}

\begin{figure}[H]
\centering
\includegraphics[width=\textwidth]{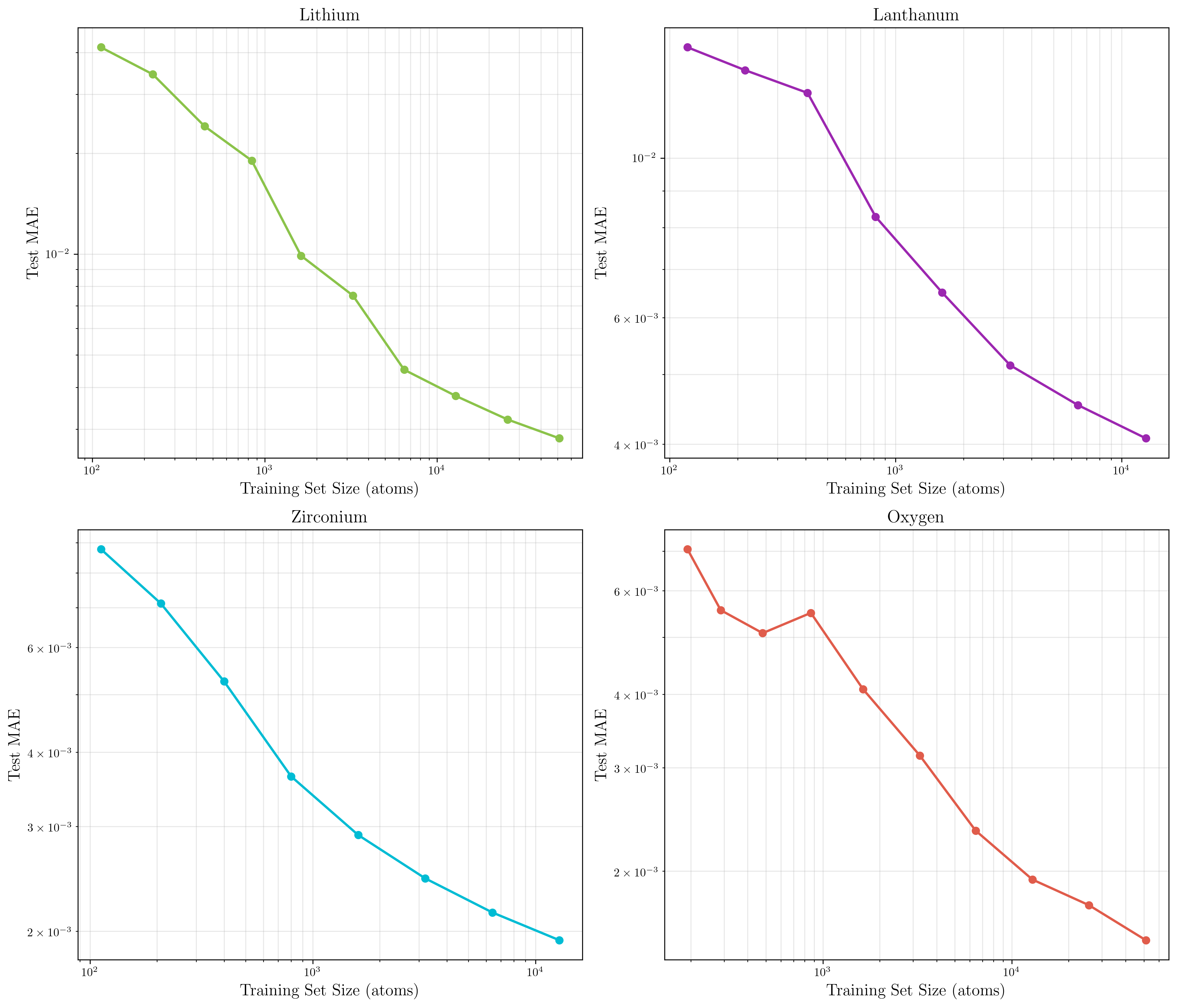}
\caption{Learning curves for LLZO DFT models (Li, La, Zr, O).}
\label{fig:lc_llzo_dft}
\end{figure}

\begin{figure}[H]
\centering
\includegraphics[width=\textwidth]{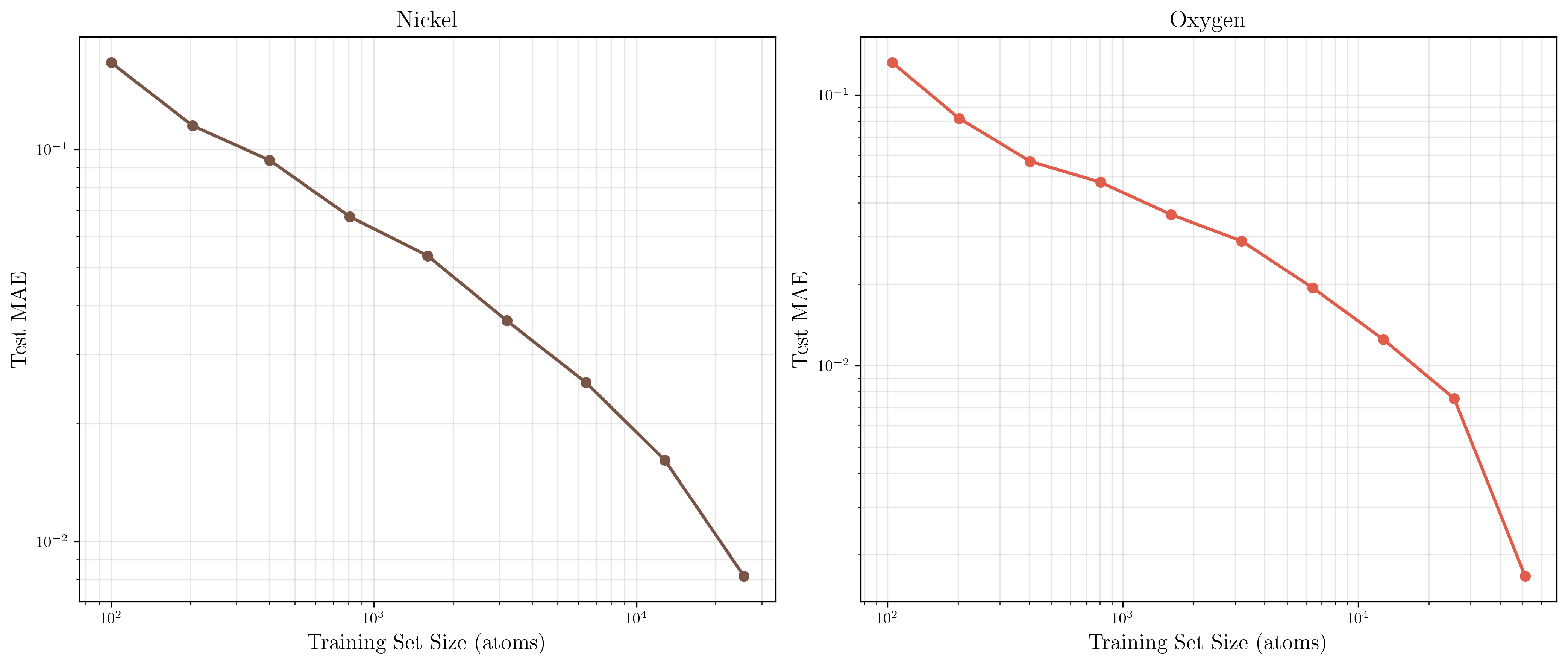}
\caption{Learning curves for Ni$_x$O$_y$ DFT models (Ni, O).}
\label{fig:lc_nio_dft}
\end{figure}

\section{SCC Cycle Reduction}

\begin{figure}[H]
\centering
\includegraphics[width=\textwidth]{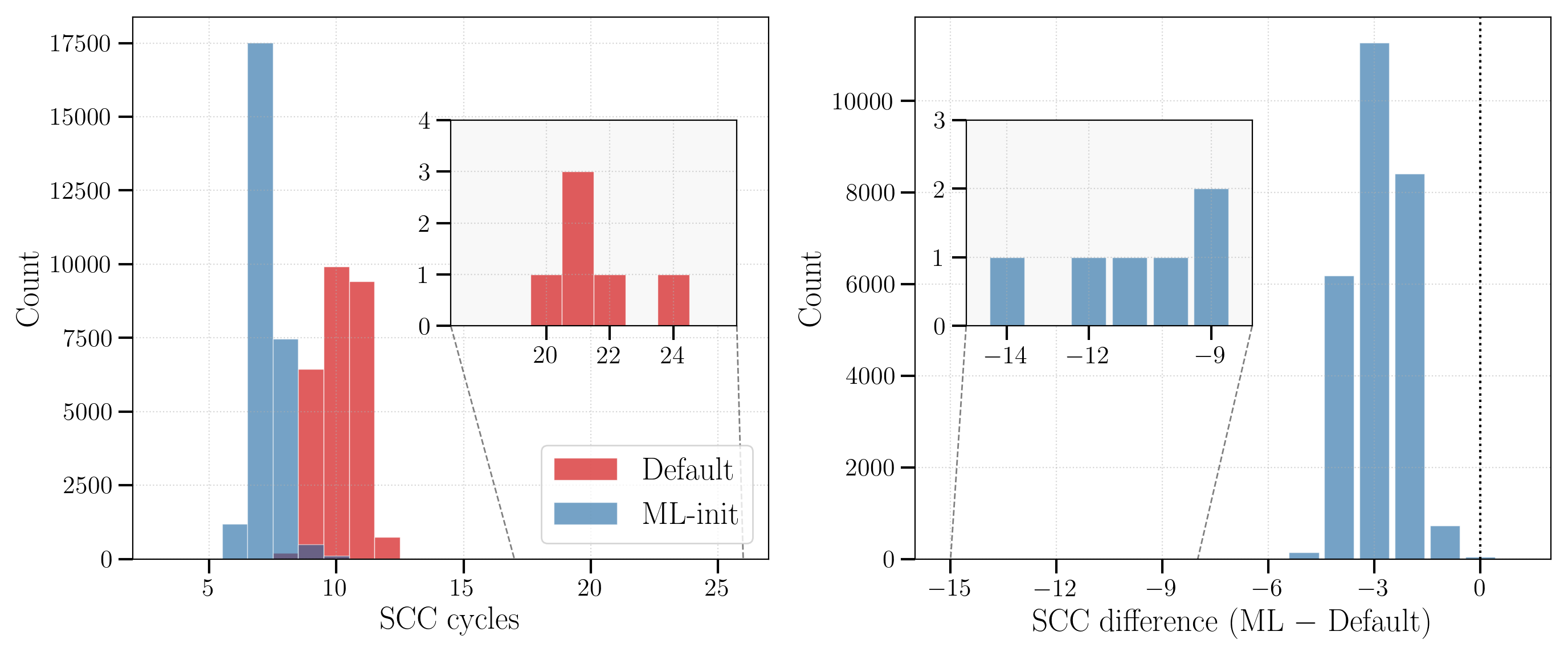}
\caption{SCC cycle distribution (left) and reduction in SCC cycles (right) for ML-initialised vs.\ standard DFTB calculations on QM9 (3ob parametrization).}
\label{fig:scc_qm9}
\end{figure}

\begin{figure}[H]
\centering
\includegraphics[width=\textwidth]{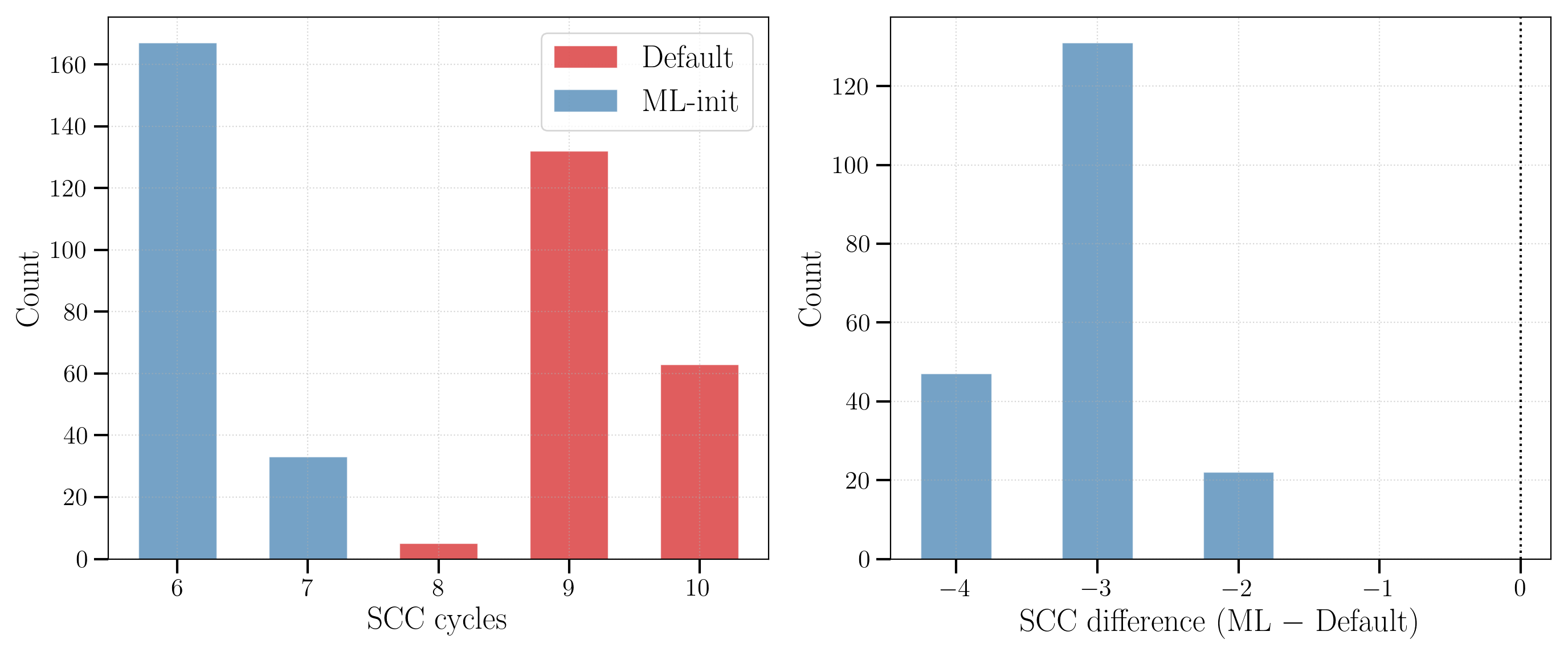}
\caption{SCC cycle distribution (left) and reduction in SCC cycles (right) for ML-initialised vs.\ standard DFTB calculations on water clusters (3ob parametrization).}
\label{fig:scc_water}
\end{figure}

\begin{figure}[H]
\centering
\includegraphics[width=\textwidth]{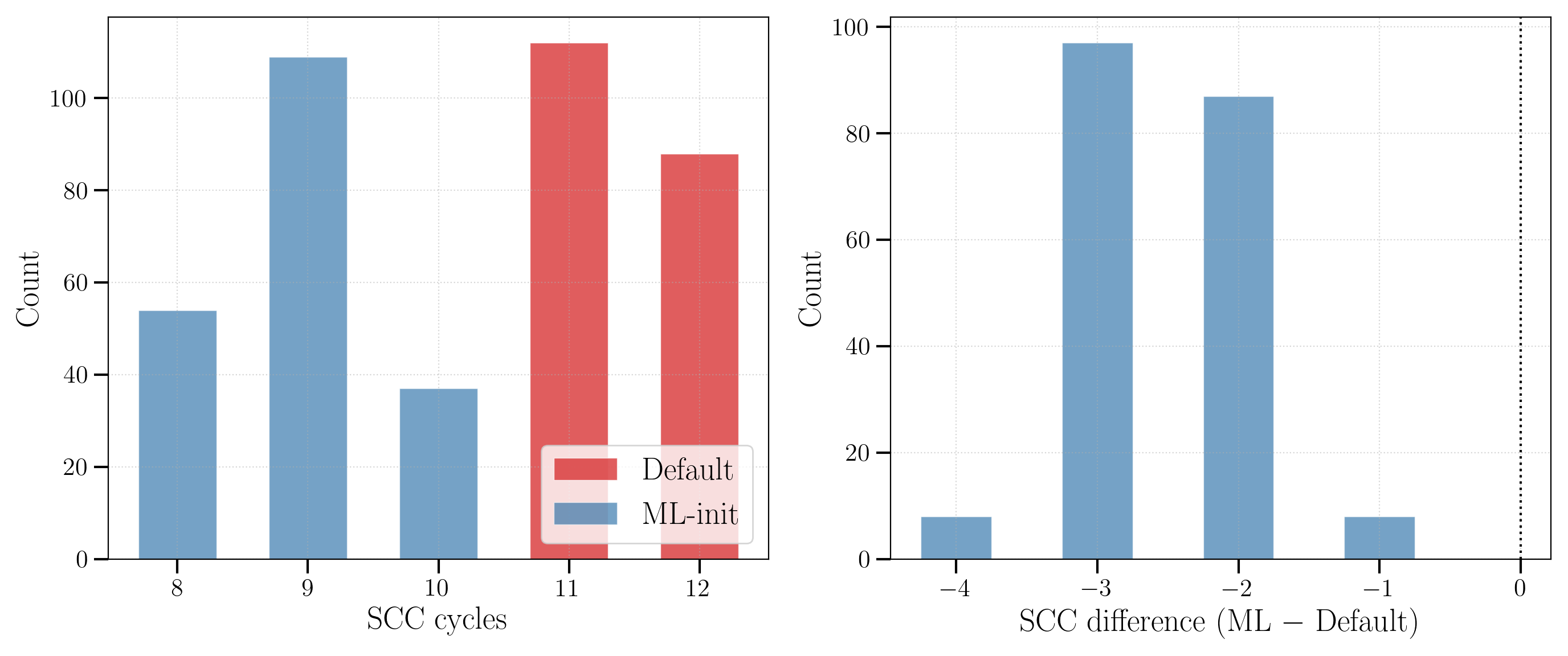}
\caption{SCC cycle distribution (left) and reduction in SCC cycles (right) for ML-initialised vs.\ standard DFTB calculations on solvated amino acids (3ob parametrization).}
\label{fig:scc_solaa}
\end{figure}

\begin{figure}[H]
\centering
\includegraphics[width=\textwidth]{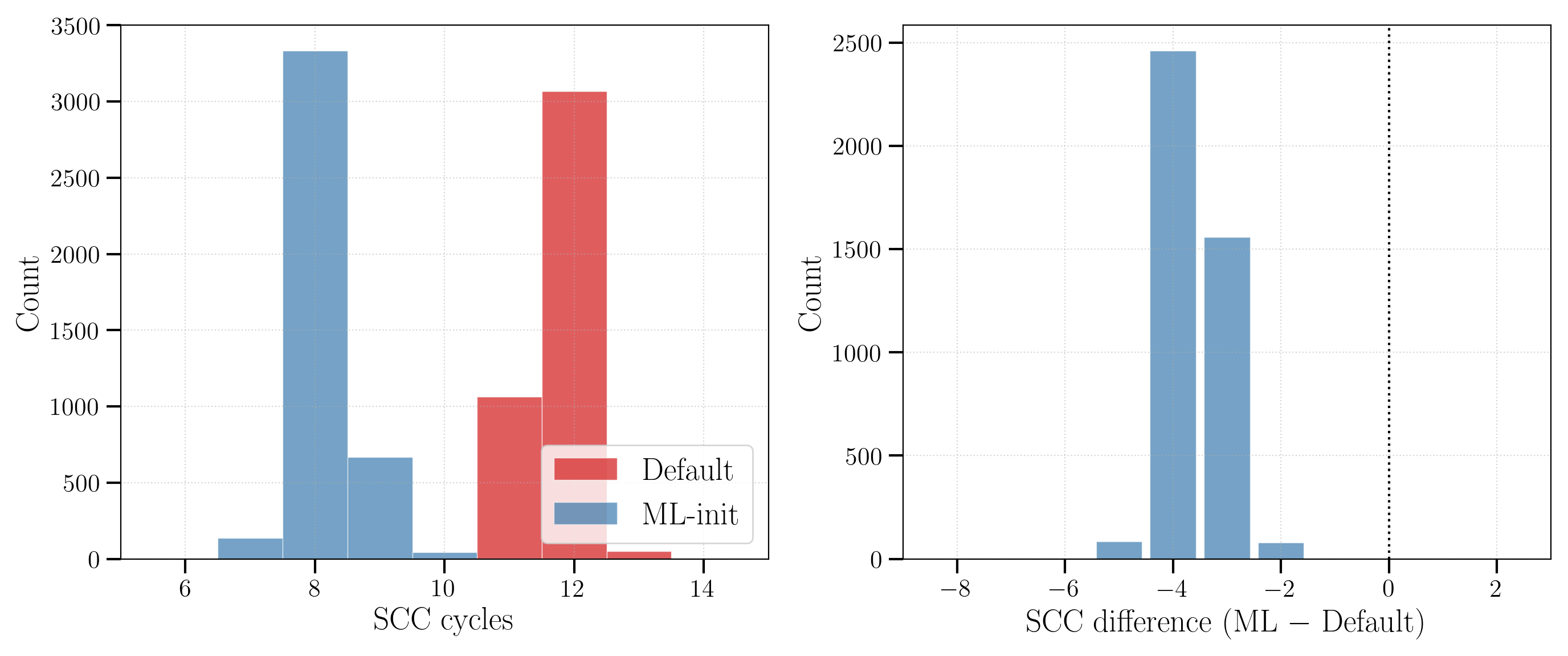}
\caption{SCC cycle distribution (left) and reduction in SCC cycles (right) for ML-initialised vs.\ standard DFTB calculations on dipeptides (3ob parametrization).}
\label{fig:scc_dipep}
\end{figure}

\begin{figure}[H]
\centering
\includegraphics[width=\textwidth]{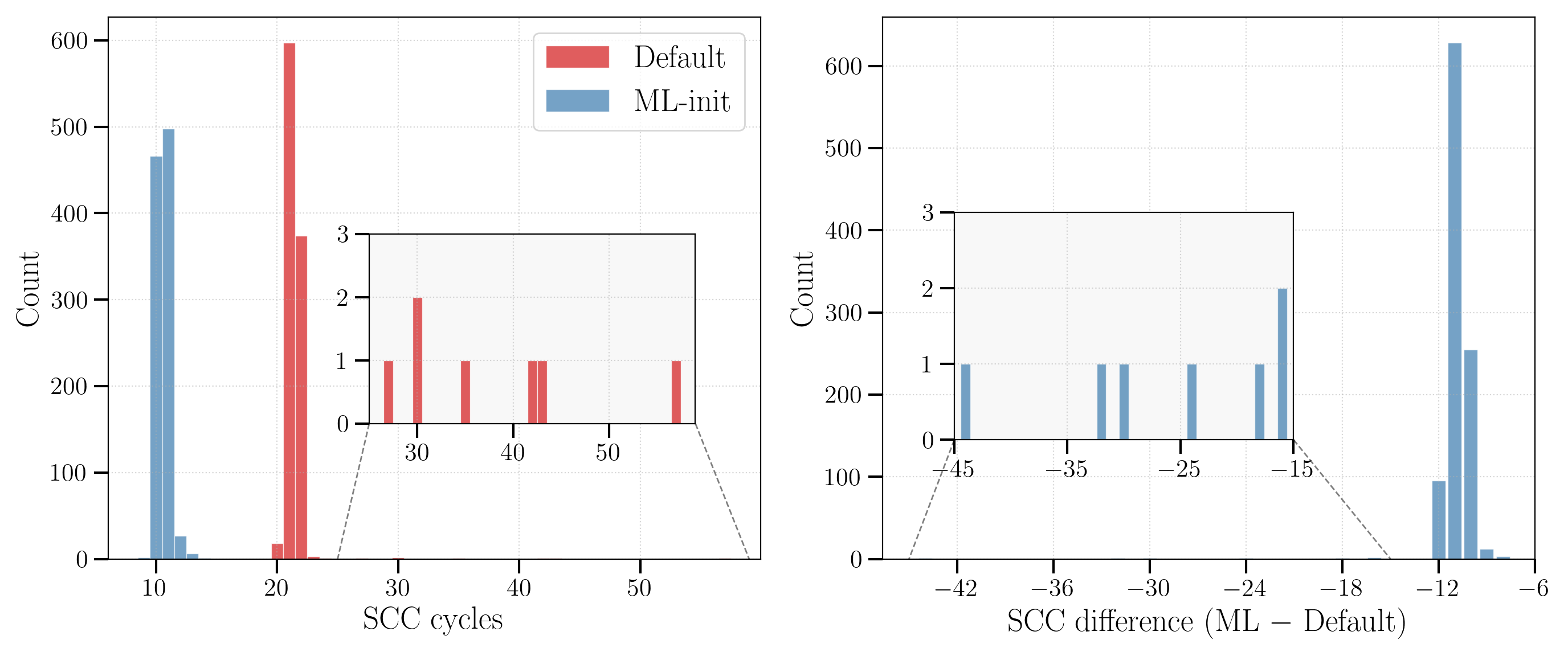}
\caption{SCC cycle distribution (left) and reduction in SCC cycles (right) for ML-initialised vs.\ standard DFTB calculations on LLZO (PTBP parametrization).}
\label{fig:scc_llzo}
\end{figure}

\section{Charge Comparison Across DFTB parametrizations}

\begin{figure}[H]
    \centering
    \includegraphics[width=\textwidth]{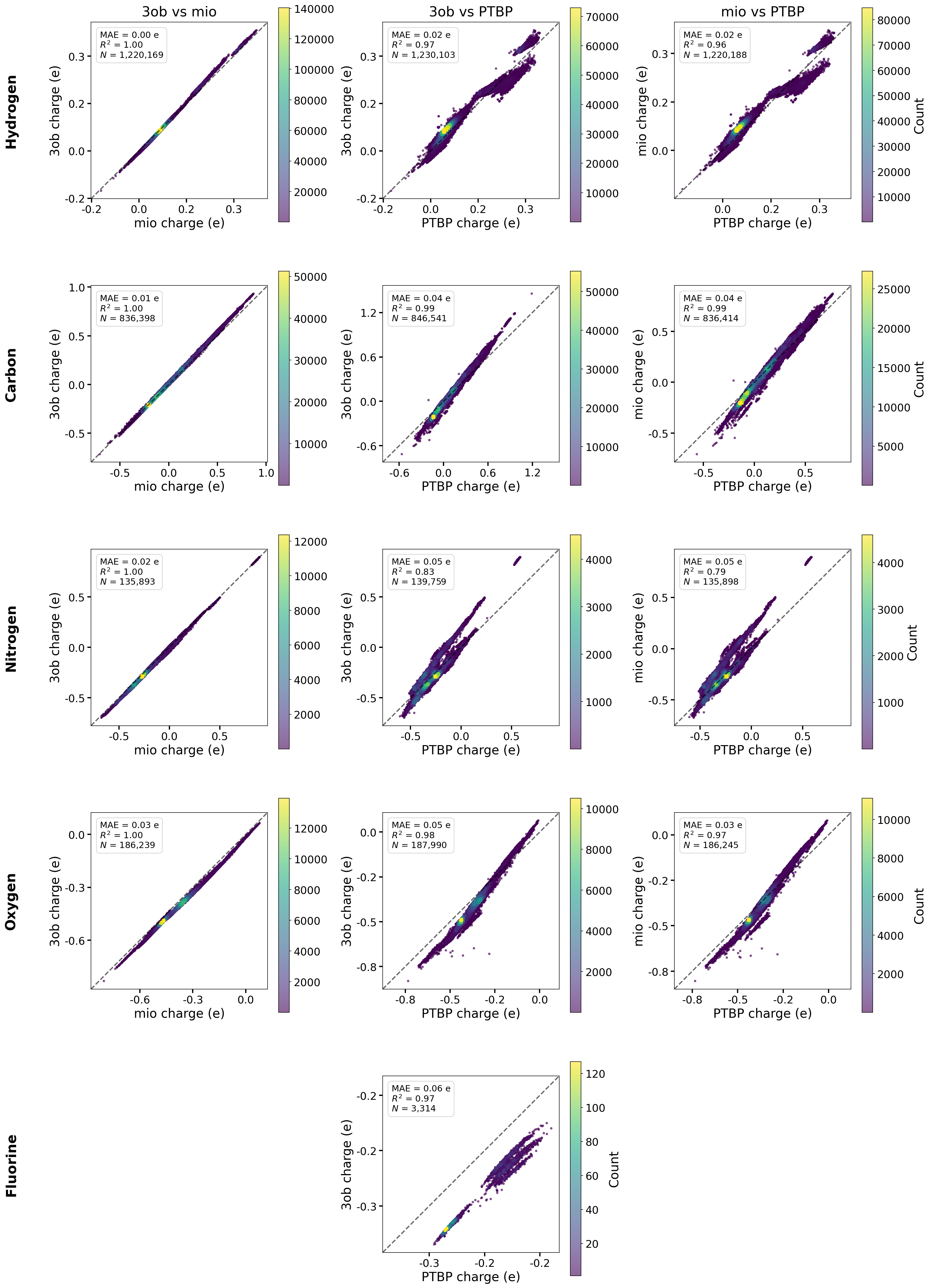}
    \caption{Pairwise comparison of DFTB Mulliken charges for the three parametrizations (3ob, mio, PTBP) on the QM9 dataset. Rows correspond to elements (H, C, N, O, F) and columns to parametrization pairs. Fluorine is not included in the mio parametrization. Color indicates point density.}
    \label{fig:dftb_params_comparison_qm9}
\end{figure}

\begin{figure}[H]
    \centering
    \includegraphics[width=\textwidth]{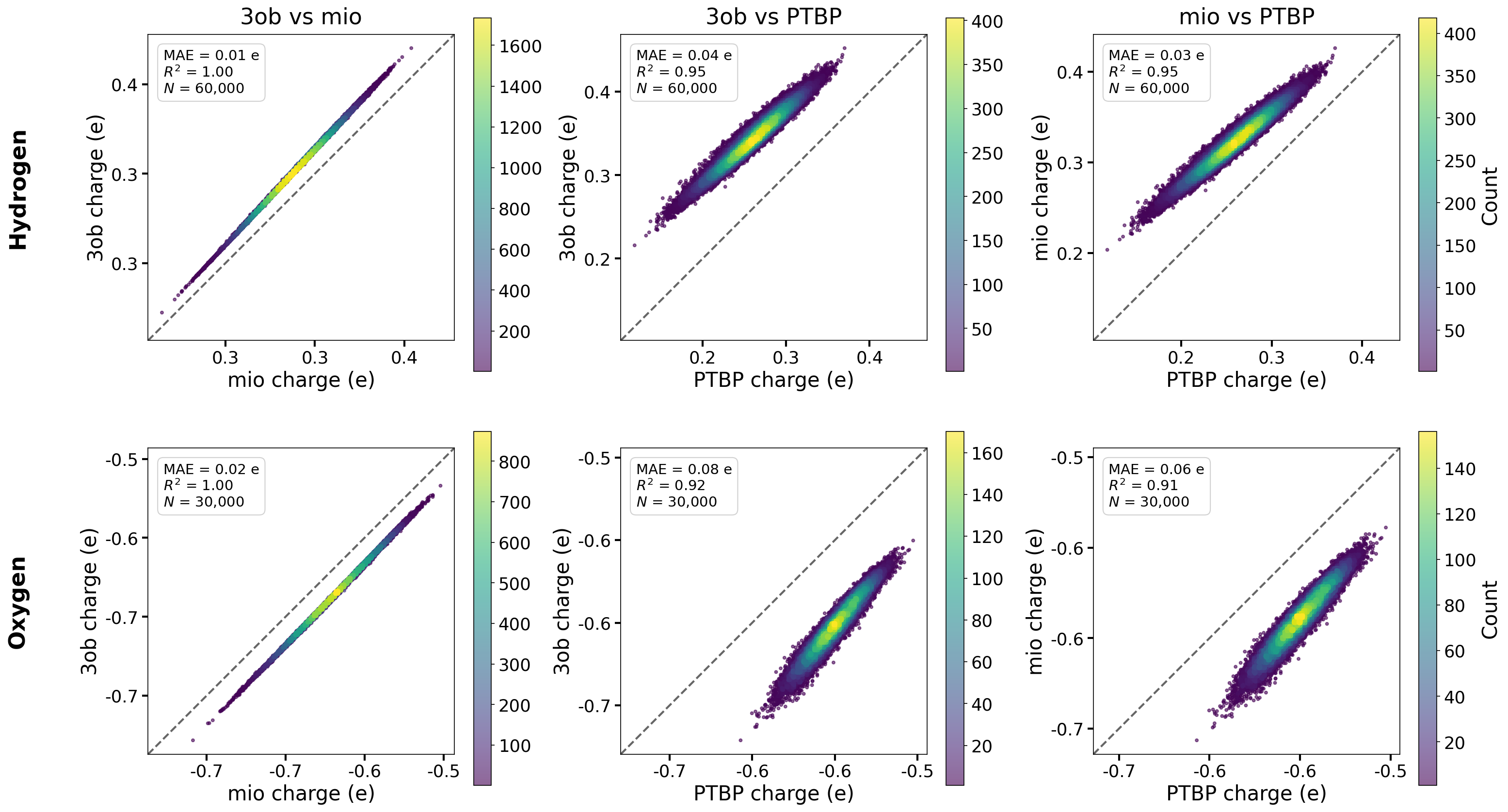}
    \caption{Pairwise comparison of DFTB Mulliken charges for the three parametrizations (3ob, mio, PTBP) on the water clusters dataset. Rows correspond to elements (H, O) and columns to parametrization pairs. Color indicates point density.}
    \label{fig:dftb_params_comparison_water}
\end{figure}

\begin{figure}[H]
    \centering
    \includegraphics[width=\textwidth]{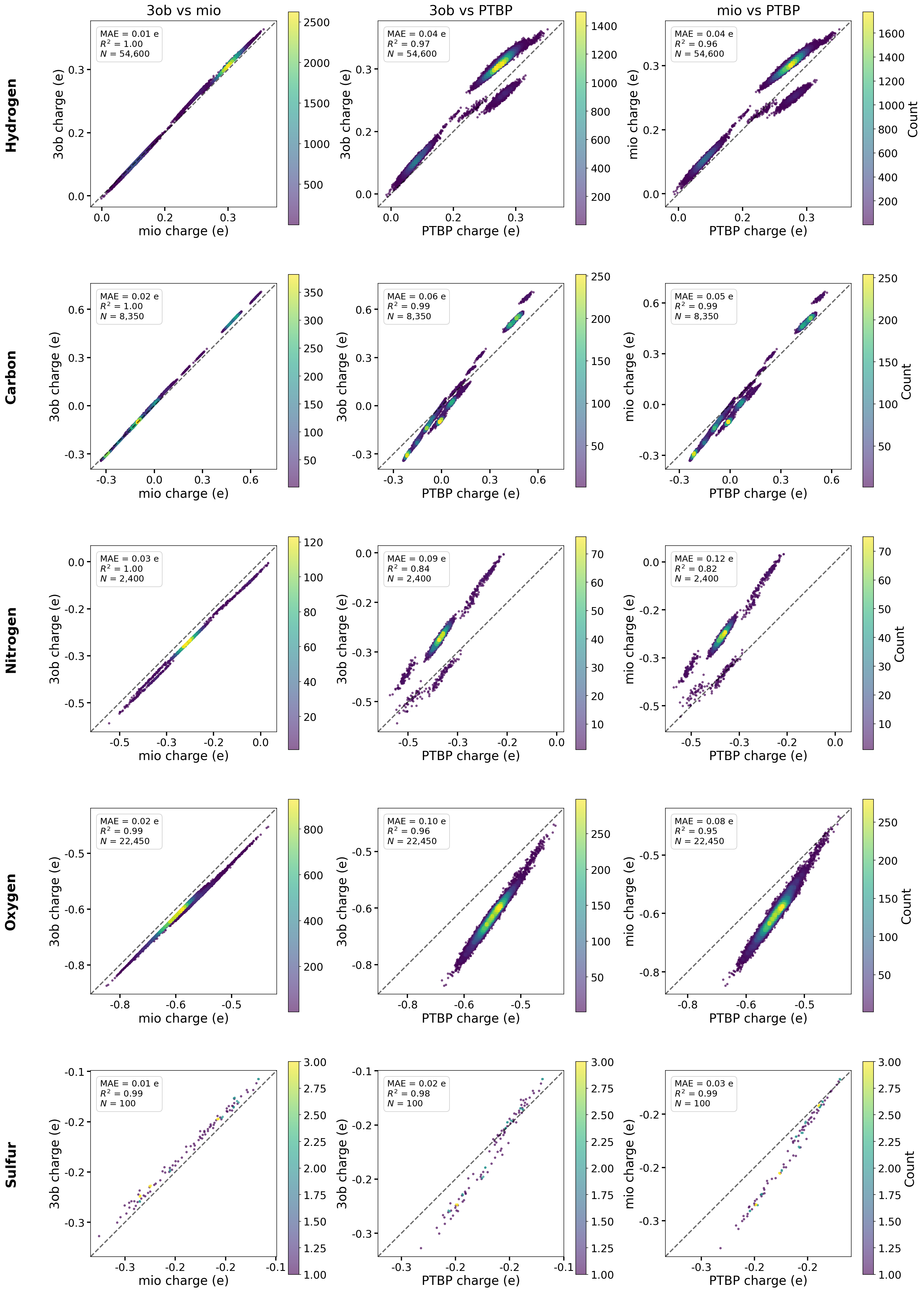}
    \caption{Pairwise comparison of DFTB Mulliken charges for the three parametrizations (3ob, mio, PTBP) on the solvated amino acids dataset. Rows correspond to elements (H, C, N, O, S) and columns to parametrization pairs. Color indicates point density.}
    \label{fig:dftb_params_comparison_solaa}
\end{figure}

\begin{figure}[H]
    \centering
    \includegraphics[width=\textwidth]{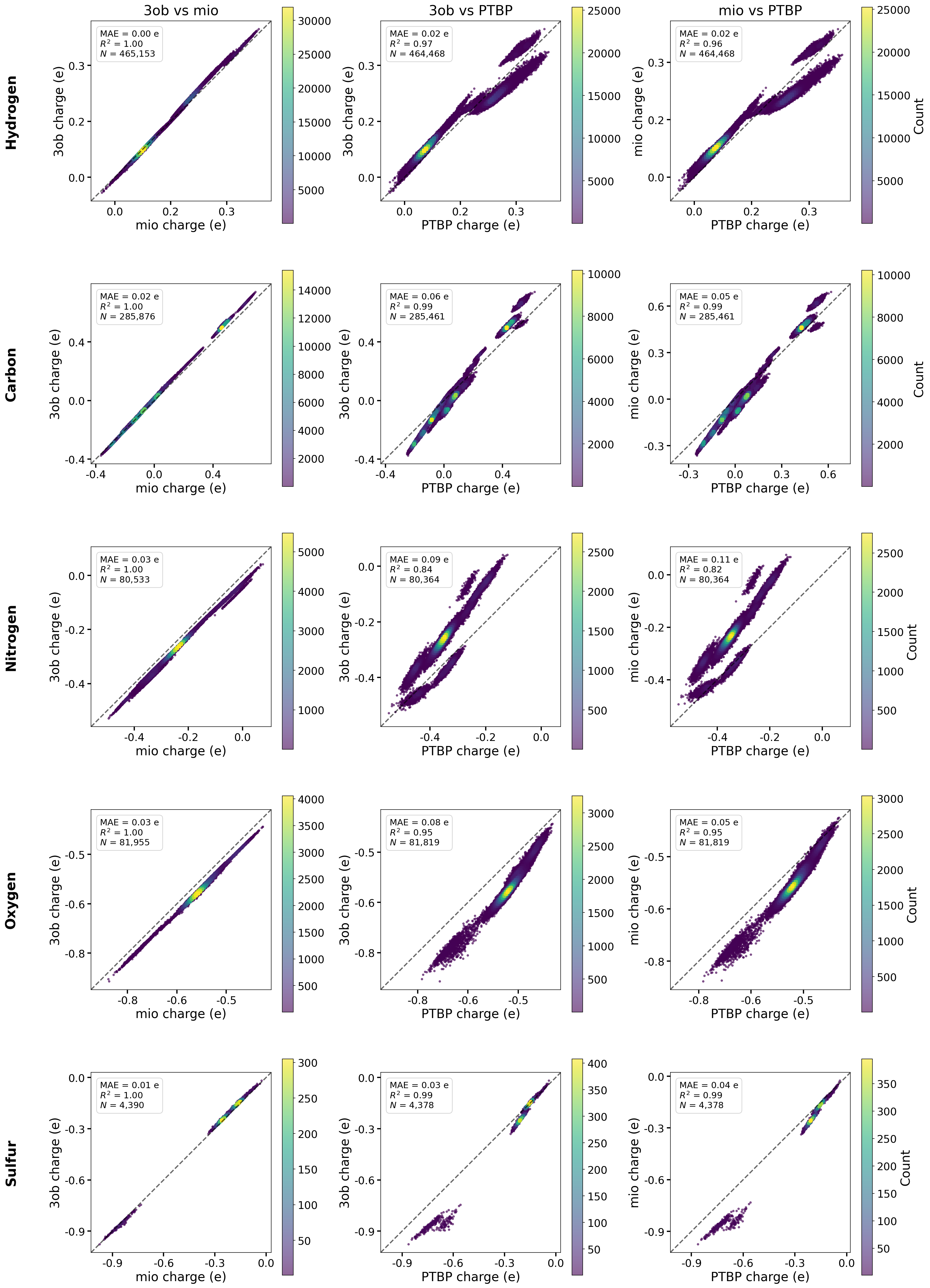}
    \caption{Pairwise comparison of DFTB Mulliken charges for the three parametrizations (3ob, mio, PTBP) on the dipeptides dataset. Rows correspond to elements (H, C, N, O, S) and columns to parametrization pairs. Color indicates point density.}
    \label{fig:dftb_params_comparison_dipep}
\end{figure}

\section{DFTB vs.\ DFT Charge Comparisons}
\label{sec:dft_vs_dftb}

\begin{figure}[H]
\centering
\includegraphics[width=\textwidth]{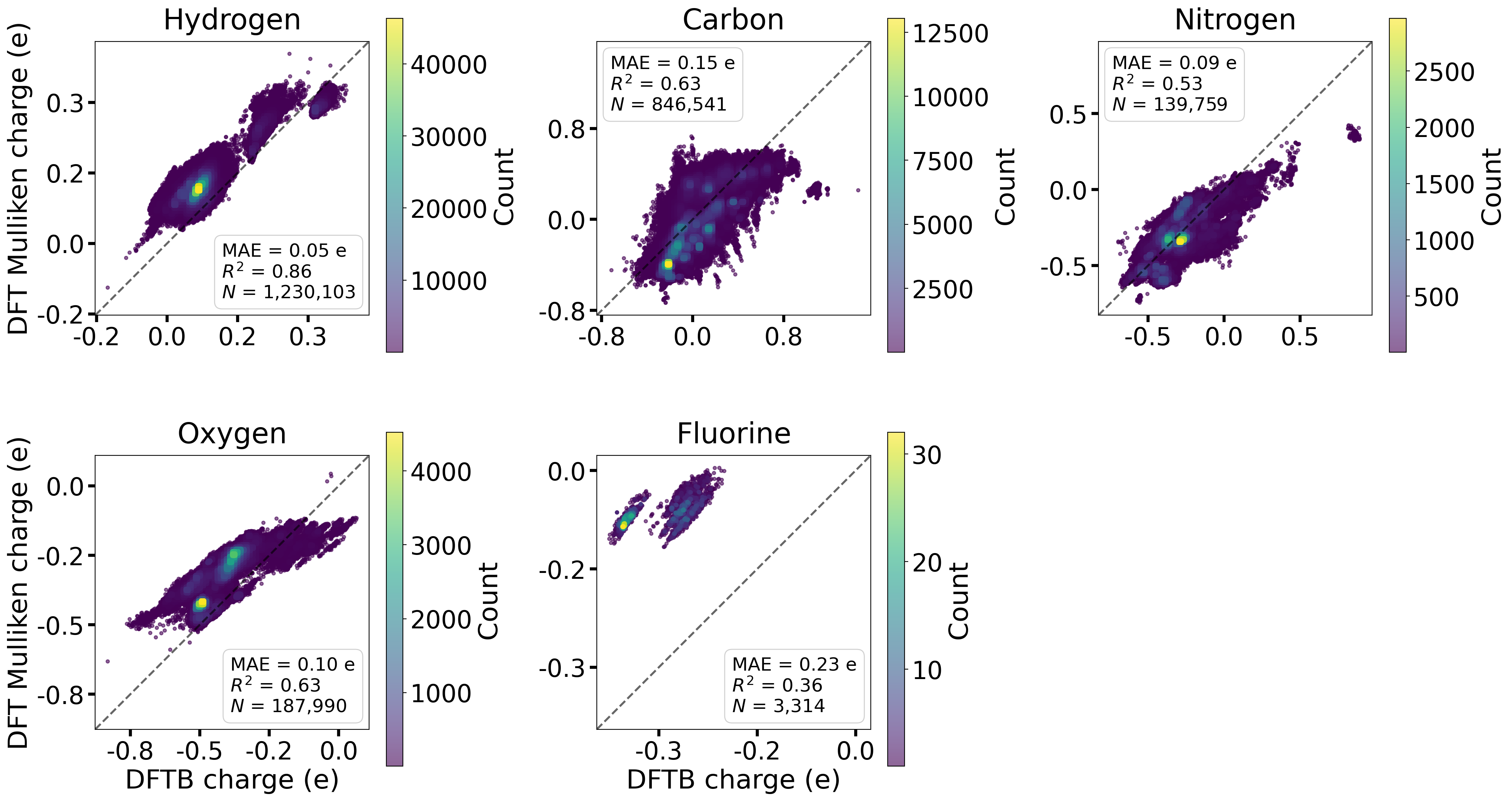}
\caption{Parity plots comparing DFT Mulliken charges vs.\ DFTB Mulliken charges for QM9 (H, C, N, O, F) using the 3ob parametrization. Color indicates point density.}
\label{fig:dft_dftb_qm9}
\end{figure}

\begin{figure}[H]
\centering
\includegraphics[width=\textwidth]{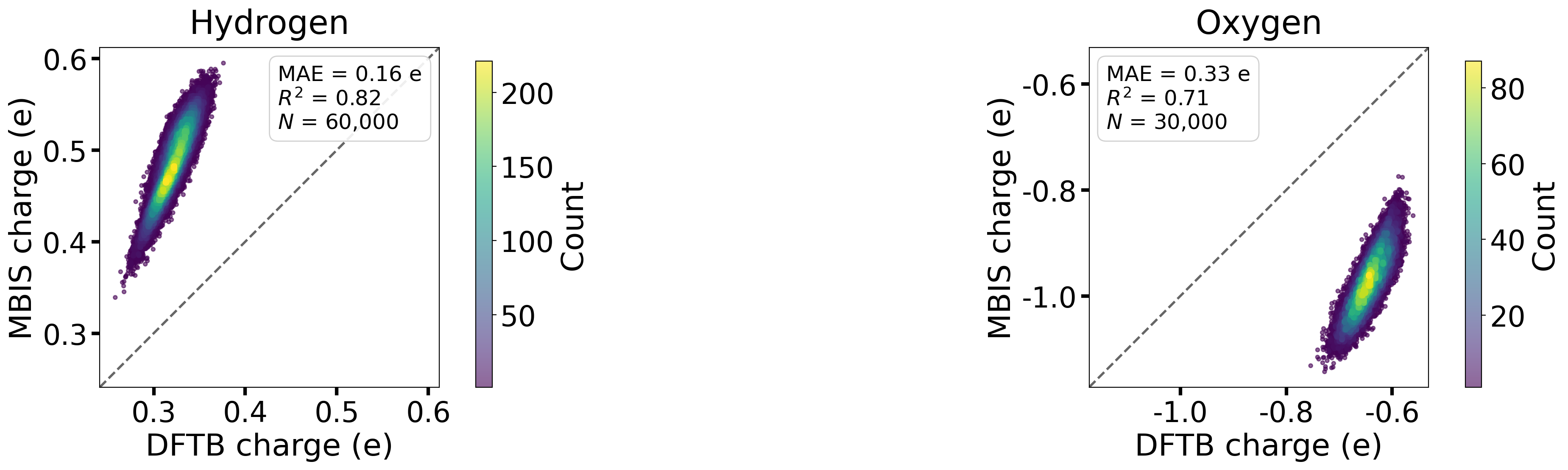}
\caption{Parity plots comparing MBIS charges vs.\ DFTB Mulliken charges for water clusters (H, O).}
\label{fig:mbis_dftb_water}
\end{figure}

\begin{figure}[H]
\centering
\includegraphics[width=\textwidth]{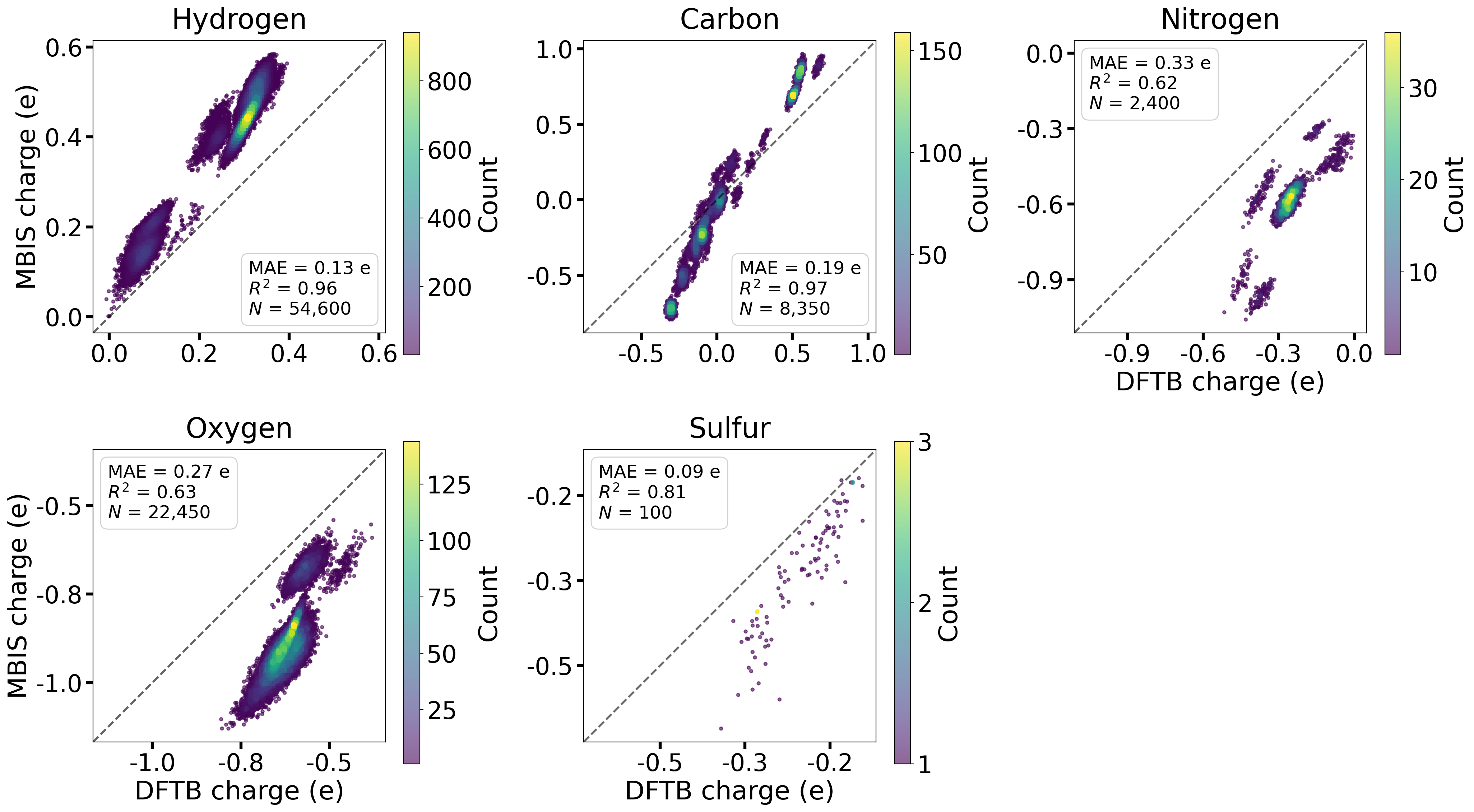}
\caption{Parity plots comparing MBIS charges vs.\ DFTB Mulliken charges for solvated amino acids (H, C, N, O, S).}
\label{fig:mbis_dftb_solaa}
\end{figure}

\begin{figure}[H]
\centering
\includegraphics[width=\textwidth]{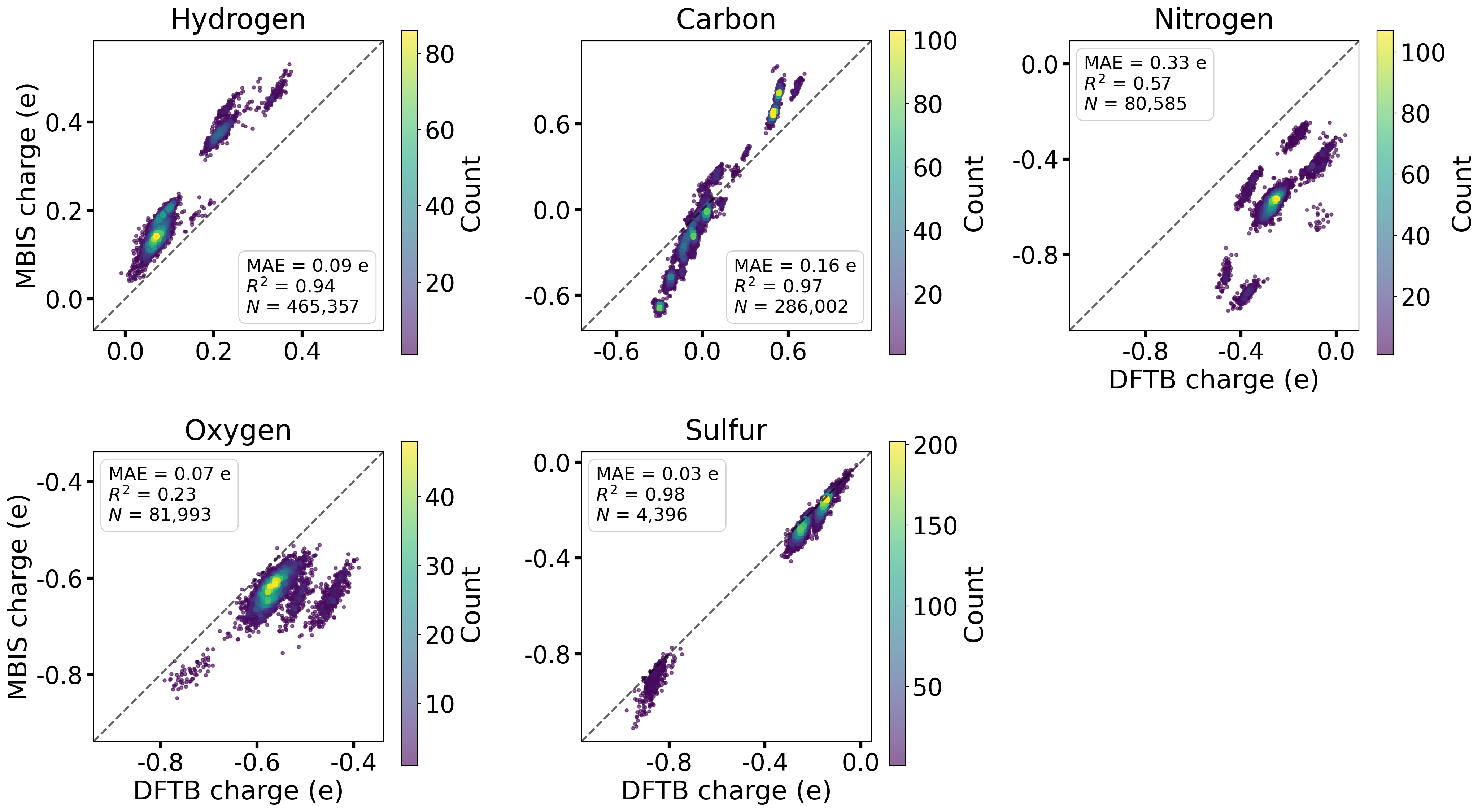}
\caption{Parity plots comparing MBIS charges vs.\ DFTB Mulliken charges for dipeptides (H, C, N, O, S).}
\label{fig:mbis_dftb_dipeptides}
\end{figure}

\begin{figure}[H]
\centering
\includegraphics[width=\textwidth]{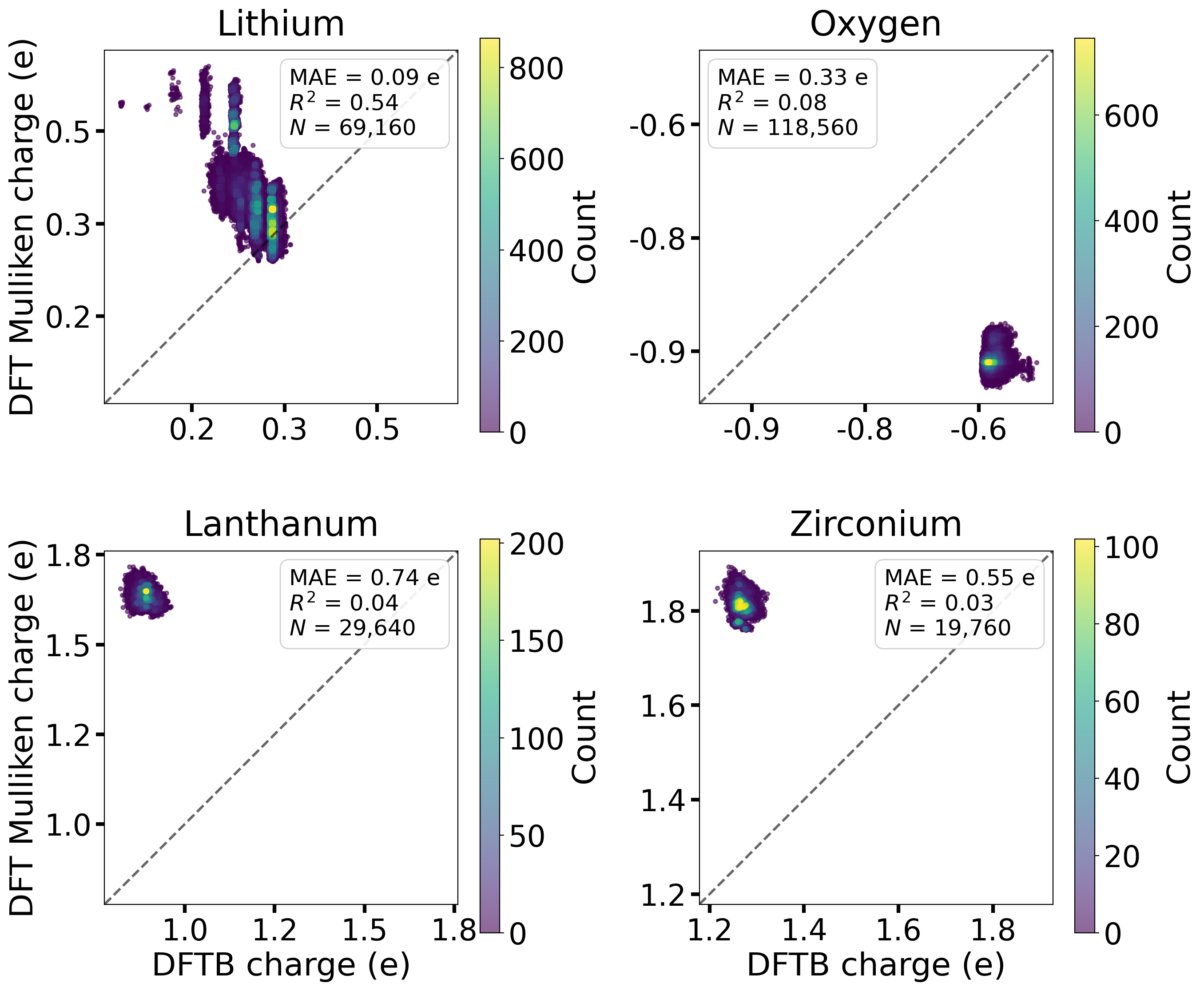}
\caption{Parity plots comparing DFT Mulliken charges vs.\ DFTB Mulliken charges for LLZO (Li, La, Zr, O) using the PTBP parametrization.
DFTB charges were recomputed on all 1,235 structures for which DFT reference data is available, using the same calculation parameters as for the random training set.
The slightly visible stripe pattern along the DFT axis arises from the limited three-decimal-place precision of the Mulliken charges in the ONETEP output, discretising the DFT charge axis at 0.001\,e intervals (cf. Fig. \ref{fig:parity_llzo_dft}).}
\label{fig:dft_dftb_llzo}
\end{figure}

\begin{figure}[H]
\centering
\includegraphics[width=\textwidth]{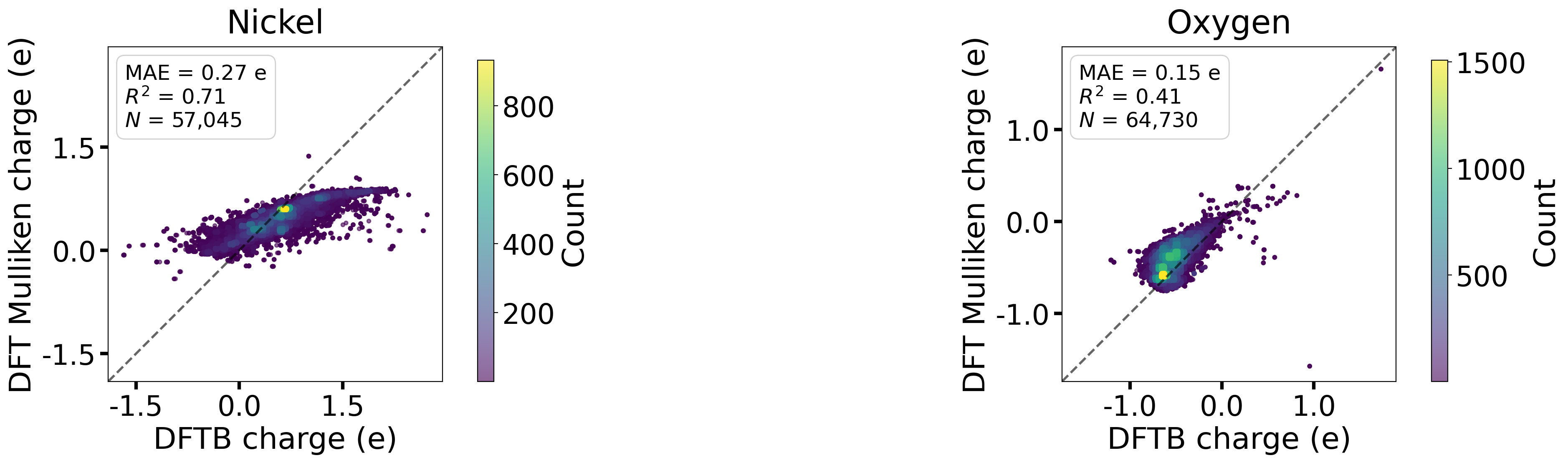}
\caption{Parity plots comparing DFT Mulliken charges vs.\ DFTB Mulliken charges for Ni$_x$O$_y$ (Ni, O) using the mio/trans3d parametrization.}
\label{fig:dft_dftb_nio}
\end{figure}

\clearpage
\section{SCC Convergence with Stricter Threshold}

\begin{table*}[htbp]
\centering
\caption{SCC convergence performance for different charge initialization strategies across all datasets, using a tighter SCC convergence criterion of $10^{-6}$\,e.
Reduction and fewer/equal/more percentages are relative to default zero-charge initialization.}
\label{tab:scc_convergence_strict}
\resizebox{\linewidth}{!}{%
\begin{tabular}{clccccc}
\toprule
Dataset & Initialization & SCC Cycles & Reduction (\%) & Fewer Cycles (\%) & Equal Cycles (\%) & More Cycles (\%) \\
\midrule
\multirow{4}{*}{QM9}
 & Default  & 11.6 & ---          & ---          & ---          & ---          \\ 
 & DFTB-ML       & 8.5 & 26.3 & 99.7 & 0.3 & 0.0 \\ 
 & Gasteiger     & 10.5 & 8.7 & 75.7 & 22.0 & 2.3 \\ 
 & DFT charges   & 10.5 & 9.1 & 80.0 & 18.7 & 1.3 \\ 
\midrule
\multirow{4}{*}{Water clusters}
 & Default  & 10.1 & ---          & ---          & ---          & ---          \\ 
 & DFTB-ML       & 7.1 & 30.3 & 100.0 & 0.0 & 0.0 \\ 
 & Gasteiger     & 9.0 & 11.0 & 99.5 & 0.5 & 0.0 \\ 
 & DFT charges   & 9.6 & 5.5 & 50.0 & 50.0 & 0.0 \\ 
\midrule
\multirow{4}{*}{Solvated amino acids}
 & Default  & 13.4 & ---          & ---          & ---          & ---          \\ 
 & DFTB-ML       & 10.7 & 19.9 & 100.0 & 0.0 & 0.0 \\ 
 & Gasteiger     & 12.5 & 7.2 & 85.5 & 14.5 & 0.0 \\ 
 & DFT charges   & 12.4 & 7.6 & 87.5 & 12.0 & 0.5 \\ 
\midrule
\multirow{4}{*}{Dipeptides}
 & Default  & 13.9 & ---          & ---          & ---          & ---          \\ 
 & DFTB-ML       & 10.0 & 28.0 & 100.0 & 0.0 & 0.0 \\ 
 & Gasteiger     & 12.7 & 8.5 & 83.4 & 16.6 &  0.0\\ 
 & DFT charges   & 12.4 & 10.7 & 91.4 & 8.6 & 0.0 \\ 
 \midrule
 \multirow{4}{*}{LLZO}
  & Default  & 21.5 & ---          & ---          & ---          & ---          \\ 
  & DFTB-ML       & 10.6 & 50.8 & 100.0 & 0.0 & 0.0 \\ 
  & Gasteiger     & --- & --- & --- & --- & --- \\
  & DFT charges   & 20.7 & 3.9 & 62.2 & 37.5 & 0.3 \\ 
 \midrule
\multirow{4}{*}{Ni$_x$O$_y$}
 & Default  & 64.2 & ---          & ---          & ---          & ---          \\ 
 & DFTB-ML       & 11.6 & 81.9 & 99.4 & 0.2 & 0.4 \\ 
 & Gasteiger     & --- & --- & --- & --- & --- \\
 & DFT charges   & 24.1 & 62.5 & 85.5 & 5.7 & 8.8 \\ 
\bottomrule
\end{tabular}}
\end{table*}

\clearpage
\section{DFTB3 SCC Convergence}

\begin{table*}[htbp]
\centering
\caption{SCC convergence performance for DFTB3-ML charge initialization across QM9 and SPICE datasets (3ob parametrization).
Reduction and fewer/equal/more percentages are relative to default zero-charge initialization.}
\label{tab:scc_convergence_dftb3}
\resizebox{\linewidth}{!}{%
\begin{tabular}{clccccc}
\toprule
Dataset & Initialization & SCC Cycles & Reduction (\%) & Fewer Cycles (\%) & Equal Cycles (\%) & More Cycles (\%) \\
\midrule
\multirow{2}{*}{QM9}
 & Default   & 10.3 & ---          & ---          & ---          & ---          \\ 
 & DFTB3-ML  & 7.3 & 29.6 & 99.9 & 0.1 & 0.0 \\ 
\midrule
\multirow{2}{*}{Water clusters}
 & Default   & 9.3 & ---          & ---          & ---          & ---          \\ 
 & DFTB3-ML  & 6.0 & 35.3 & 100.0 & 0.0 & 0.0 \\ 
\midrule
\multirow{2}{*}{Solvated amino acids}
 & Default   & 11.4 & ---          & ---          & ---          & ---          \\ 
 & DFTB3-ML  & 8.8 & 23.1 & 100.0 & 0.0 & 0.0 \\ 
\midrule
\multirow{2}{*}{Dipeptides}
 & Default   & 11.2 & ---          & ---          & ---          & ---          \\ 
 & DFTB3-ML  & 8.0 & 28.7 & 100.0 & 0.0 & 0.0 \\ 
\bottomrule
\end{tabular}}
\end{table*}

\clearpage
\section{Cross-Parameterization Transferability}

\begin{table*}[htbp]
\centering
\caption{Cross-parameterization SCC convergence: mean SCC cycle counts for all combinations of DFTB parameterization (rows) and charge model parameterization (columns) across four molecular datasets.
Values in parentheses are percentage reductions relative to the default zero-charge initialization for the same DFTB parameterization.
Diagonal entries, where the charge model and DFTB parameterization match, are shown in bold.
All comparisons are restricted to the matched testset (subset) common to each charge model and its corresponding vanilla calculation.
The mio parametrization does not include fluorine, so fluorine-containing molecules are excluded from both the mio charge models and the corresponding QM9 test sets.}
\label{tab:cross_param}
\resizebox{\linewidth}{!}{%
\begin{tabular}{llccccc}
\toprule
Dataset & DFTB param. & Default & 3ob model & mio model & PTBP model \\
\midrule
\multirow{3}{*}{QM9}
 & 3ob  & 10.2 & \textbf{7.3 (28.3\%)} & 7.6 (25.3\%) & 8.3 (18.2\%) \\ 
 & mio  & 10.1 & 7.6 (24.8\%) & \textbf{7.2 (28.0\%)} & 8.1 (19.2\%) \\ 
 & PTBP & 11.3 & 9.1 (18.9\%) & 9.0 (20.3\%) & \textbf{7.9 (30.3\%)} \\ 
\midrule
\multirow{3}{*}{Water clusters}
 & 3ob  & 9.3 & \textbf{6.2 (33.6\%)} & 7.0 (25.0\%) & 8.0 (13.9\%) \\ 
 & mio  & 9.0 & 6.9 (23.0\%) & \textbf{6.1 (31.8\%)} & 8.0 (10.8\%) \\ 
 & PTBP & 10.0 & 8.8 (12.0\%) & 8.3 (16.6\%) & \textbf{7.0 (29.9\%)} \\ 
\midrule
\multirow{3}{*}{Solvated amino acids}
 & 3ob  & 11.4 & \textbf{8.9 (22.1\%)} & 9.4 (18.3\%) & 10.0 (12.8\%) \\ 
 & mio  & 11.4 & 9.4 (17.6\%) & \textbf{8.9 (22.0\%)} & 10.0 (12.6\%) \\ 
 & PTBP & 13.3 & 11.1 (16.8\%) & 11.1 (16.7\%) & \textbf{9.7 (26.8\%)} \\ 
\midrule
\multirow{3}{*}{Dipeptides}
 & 3ob  & 11.8 & \textbf{8.2 (30.8\%)} & 9.1 (22.6\%) & 9.8 (16.5\%) \\ 
 & mio  & 11.7 & 9.1 (22.3\%) & \textbf{8.1 (30.7\%)} & 9.9 (15.9\%) \\ 
 & PTBP & 13.3 & 11.5 (13.1\%) & 12.4 (6.8\%) & \textbf{9.1 (31.4\%)} \\ 
\bottomrule
\end{tabular}}
\end{table*}

\newpage
\bibliographystyle{unsrt}
\bibliography{bibliography}